\documentclass[aps,preprint,nofootinbib,endfloats,showpacs]{revtex4}

\usepackage{graphics}
\usepackage{epsfig}

\begin{document}

\title{The Fermion Monte Carlo revisited}

\author{Roland Assaraf$^1$, Michel Caffarel$^2$, and Anatole Khelif$^3$}
\affiliation{
$^{1}$Laboratoire de Chimie Th{\'e}orique CNRS-UMR 7616, Universit{\'e} Pierre 
et Marie Curie, 4 Place Jussieu, 75252 Paris, France \\
$^{2}$Laboratoire de Chimie et Physique Quantiques CNRS-UMR 5626, IRSAMC, Universit\'e Paul 
Sabatier, 118 route de Narbonne 31062 Toulouse Cedex, France \\
$^{3}$Laboratoire de Logique Math\'ematique, Universit{\'e} Denis Diderot, 
2 Place Jussieu, 75251 Paris, France}

\date{\today}

\begin{abstract}
In this work we present a detailed study of the Fermion Monte Carlo algorithm
(FMC), a recently proposed stochastic method for calculating fermionic
ground-state energies. A proof that the FMC method is an exact method is given.
In this work the stability of the method is related to the difference
between the lowest (bosonic-type) eigenvalue of the FMC diffusion operator and the
exact fermi energy. It is shown that within a FMC framework
the lowest eigenvalue of the new diffusion operator is no longer the bosonic 
ground-state eigenvalue as in standard exact Diffusion Monte Carlo (DMC) schemes 
but a modified value which is strictly greater.
Accordingly, FMC can be viewed as an exact DMC method
built from a correlated diffusion process having a {\it reduced} Bose-Fermi gap.
As a consequence, the FMC method is more stable than any transient method (or
nodal release-type approaches). It is shown that the most recent ingredient
of the FMC approach [M.H. Kalos and F. Pederiva, Phys. Rev. Lett.
{\bf 85}, 3547 (2000)], namely the introduction of non-symmetric guiding
functions, does not necessarily improve the stability of the algorithm. 
We argue that the stability observed with such
guiding functions is in general a finite-size population effect disappearing for a 
very large population of walkers. 
The counterpart of this stability is a control population error which is 
different in nature from the standard Diffusion Monte Carlo algorithm
and which is at the origin of an uncontrolled approximation in FMC.
We illustrate the various ideas presented in this work with calculations
performed on a very simple model having only nine states but a full ``sign problem''. 
Already for this toy model it is clearly seen that FMC calculations are inherently 
uncontrolled.
\end{abstract}

\pacs{  02.70Ss, 05.30.Fk  }
\maketitle

\section{Introduction}
In theory quantum Monte Carlo (QMC) techniques\cite{kalosfermion99} 
are capable of giving an exact estimate of the energy with an evaluation of the error: 
the statistical error. Unfortunately, such an ideal situation is not realized in practice. 
Exact results with a controlled finite statistical error are only achieved for bosonic systems.
For fermionic systems, we do not have at our disposal an algorithm which is both exact 
and stable (statistical fluctuations going to zero in the large simulation time regime).
This well-known problem is usually referred to as the ``sign problem''. 
The usual solution to cope with this difficulty consists in defining a stable algorithm 
based on an uncontrolled approximation, the so-called Fixed-Node approximation.
\cite{ceperleykalos79,schmittkalos84,lesterbook94,reynoldsJCP82,andersonJCP75,andersonJCP76}
In practice, the fixed-node error on the energy is small when one uses good trial
wavefunctions and, thus, QMC methods can be considered today as reference methods to
compute groundstate energies as shown by a large variety of 
applications\cite{reynoldsJCP82,filippiJCP96,surf1,surf2,needs1,needs2,MitasGrossman00}. 
However, the accuracy of the results is never known from the calculation, it is known only 
{\it a posteriori}, for example by comparison with experimental data.
Exact methods, which are basically transient 
methods\cite{ceperleyalderPRL84,bernuJCP90,bernuJCP91,caffarelJCP1992}
including the nodal release method\cite{ceperleyalderPRL84}, have been applied with success only
to very specific models (small or homogeneous systems) for which the sign instability 
is not too severe (small Bose-Fermi energy gap).

Recently, Kalos and coworkers \cite{kalosfermion00} have proposed a method
presented as curing the sign problem, the so-called Fermion Monte Carlo method (FMC).
This work makes use of two previously introduced ingredients, a cancellation process 
between ``positive'' and ``negative'' walkers introduced by Arnow {\sl et al.} \cite{arnow1982}
and a modified process correlating explicitly the dynamics of the walkers 
of different signs.\cite{Liu1994}
The new feature introduced in \cite{kalosfermion00} is the introduction of non-symmetric 
guiding functions. The method has been tested on various simple systems including free fermions  
and interacting systems 
such as the $^3$He fluid.\cite{kalosfermion00,CollettiFMC2005}
The results are found to be compatible with the 
assumption that the method is stable and not biased. However, this conclusion is not 
clear at all because of the presence of large error bars. 
The purpose of this work is to present a detailed  analysis of the
algorithm and a definitive answer to this assumption.

The content of this paper is as follows.
In section II we give a brief presentation of the ``sign problem''.
The sign instability in an exact DMC approach comes from the blowing
up in imaginary time of the undesirable bosonic component associated with the
lowest mathematical eigenstate of the Hamiltonian (a wavefunction which is
positive and symmetric with respect to the exchange of particles).
The fluctuations of the transient energy estimator grows like $e^{t(E_0^F-E_0^B)}$, 
where $E_0^F$ is the fermionic ground-state energy (here and in what follows, the superscript 
$F$ stands for ``Fermionic'') and $E_0^B$, the lowest mathematical eigenvalue 
of the Hamiltonian (the superscript ``B'' standing for Bosonic).
In Sec. III we briefly recall the main elements of the standard DMC
method, the basic stochastic algorithm simulating the imaginary-time
evolution of the Hamiltonian. 
In section IV, we describe the FMC method as a generalization of the DMC
method and we show that {\it FMC is an exact method}, that is, no systematic bias is introduced.
This section is made of two parts.
In a first part we introduce the notion of positive and negative walkers
to represent a signed wavefunction in DMC.
This will help us to view the FMC method as a generalization of
the DMC method, the FMC introducing two important modifications
with respect to DMC: a correlated dynamics for positive and negative walkers
and a cancellation process for such pairs whenever they meet.

In section V we study the stability of the algorithm. 
We prove that the {\it FMC method is in general not stable}, the fluctuations of the transient 
estimator of the energy  growing exponentially like $e^{t(E_0^F-\tilde{E}_0^B)}$ 
where $\tilde{E}_0^B$ is the lowest bosonic-like eigenvalue 
of the generalized diffusion process operator 
associated with FMC whose expression is given explicitly. 
It is shown that {\it FMC is more stable than the standard nodal release DMC method}
because $\tilde{E}_0^B > E_0^B$.
In section VI we illustrate our theoretical results using a toy model,
a ``minimal'' quantum system having a genuine sign problem (two coupled 
oscillators on a finite lattice). The different aspects of the FMC method are 
highlighted in this application. 
In the case of non-symmetric wavefunctions introduced recently in Ref.\cite{kalosfermion00}
it is shown that, in contrast with DMC where the population control
error decays linearly as a function of the population size, the FMC decay displays a much
more slower power law. 
As a consequence, one needs a very large
population of walkers to remove the control population error 
and to observe the instability of FMC.
Let us emphasize that observing such a subtle finite population effect 
for a genuine many-fermion system is actually impossible. Here, to illustrate 
this important point numerically, we have been led to consider a very simple system having
only a few states.
For this system, a large population of
walkers -eventually  much larger than the
dimension of the quantum Hilbert space itself- can be considered. The results obtained 
in that regime confirm our theoretical findings, in particular the fact that the stability of the 
algorithm presented elsewhere \cite{kalosfermion00} 
is only apparent. As an important conclusion, we emphasize that the FMC control
population error is an uncontrolled approximation for realistic fermion systems.
 
\section{Fermion instability in quantum Monte Carlo}
We consider a Schr\"odinger operator for a system of $N$ fermions,
\begin{equation}
H = -\frac{1}{2} {\bf \nabla}^2 + V ({\bf R})
\label{schro}
\end{equation}
where we note ${\bf R}=({\bf r}_1 \dots {\bf r}_N)$  the $3N$ coordinates of the
$N$ particles in the three dimensional space. In this expression, 
the first term is the kinetic energy (${\bf \nabla}^2$ is the Laplacian operator 
in the space of the $3N$ coordinates). The second term, $V$, is the potential.
DMC techniques are based upon the evolution of the Hamiltonian 
in imaginary time. We express this evolution using the spectral decomposition:
\begin{equation}
e^{-t(H-E_T)} = \sum_{i} e^{-t (E_i-E_T)} \mid \phi_i \rangle \langle \phi_i \mid
\label{evolution}
\end{equation}
where $\phi_i$ are the (normalized) eigenfunctions of $H$, $E_i$ are the
corresponding eigenvalues and $E_T$ is a so-called reference energy.
The fundamental property of operator (\ref{evolution}) is to filter 
out the lowest eigenstate $\phi_0$. 
To understand how it works, we consider the time evolution of a wavefunction 
$f_0 ({\bf R})$ 
\begin{equation}
\mid f_t \rangle \equiv e^{-t (H-E_T)} \mid f_0 \rangle
\label{f_tdef}
\end{equation}
and calculate the overlap with an eigenstate $\phi_i$
\begin{equation}
\langle \phi_i \mid  f_t \rangle = e^{-t(E_i-E_T)} \langle
\phi_i \mid f_0 \rangle.
\end{equation}
From this expression it is easily seen that the component on the lowest 
eigenstate $\phi_0$ is growing exponentially faster than the higher components.
In DMC methods the function $f_t$ is generated using random
walks. For $t$ large enough the lowest eigenstate 
\begin{equation}
f_t \sim e^{-t (E_0-E_T)} \phi_0
\label{ft}
\end{equation}
is produced.
This asymptotic behaviour makes DMC an accurate method for 
computing the properties of $\phi_0$, in particular the energy  $E_0$.
Unfortunately, for fermionic systems the physical groundstate $\phi_0^F$,  which is
antisymmetric with respect to the exchange of particles, is not
$\phi_0$, but some ``mathematically'' excited state because 
$\phi_0$ is positive and symmetric for a Schr\"odinger
operator (bosonic ground-state). For the sake of clarity, 
we shall denote from now on this bosonic ground-state as $\phi_0^B$ and its energy, $E_0^B$.
The necessity of extracting an exponentially small component to evaluate
$E_0^F$ or any fermionic 
property is at the origin of the fermion instability in quantum Monte Carlo.

We now give a quantitative analysis of this instability.
Since the asymptotic behaviour of $f_t$ is not useful to compute the fermionic energy, 
we consider the transient behaviour of the evolution of $f_t$.
Basically, exact methods \cite{ceperleyalderPRL84} filter out  $\phi_0^F$ by projecting 
the transient behaviour of $f_t$ on the antisymmetric space.
Introducing the antisymmetrization operator ${\cal A}$, one obtains the
physical groundstate for $t$ large enough 
\begin{equation}
{\cal A} \mid f_t \rangle \sim e^{-t(E_0^F-E_T)} \phi_0^F
\label{aft}
\end{equation}
since the components of $f_t$ over the higher antisymmetric eigenstates are
decreasing exponentially with respect to the component on $\phi_0^F$. 
In practice, the fermionic energy $E_0^F$ is calculated using an
antisymmetric function $\psi_T$ and using the fact that, at large enough $t$, one has 
\begin{equation}
E_0^F = \frac{\langle \psi_T \mid H \mid f_t \rangle}{\langle \psi_T \mid f_t
  \rangle}.
\label{E_Fproj}
\end{equation}
In the long-time regime the stochastic estimation of the R.H.S. of Eq.(\ref{E_Fproj}) 
is unstable.
In essence, the signal, the antisymmetric component of $f_t$, 
%RA: pourquoi des - et pas des, ? MC: je ne sais pas! je trouve ca plus joli. Bon!
mettons des virgules....
decreases exponentially fast with respect to $f_t$, Eq.(\ref{ft}). The signal-over-noise 
ratio behaves like $e^{-t(E_0^F-E_0^B)}$ and, thus, an exponential growth
of the fluctuations of the DMC estimator, Eq.(\ref{E_Fproj}), appears.

Now, let us give a more quantitative analysis by writing this estimator and 
evaluating the variance.
In a standard diffusion Monte Carlo calculation, the time-dependent distribution
$f_t$  is generated by a random walk over a population of walkers $\{{\bf R}_i\}$.
Formally $f_t$ is given in the calculation as an average at time $t$ over Dirac functions 
centered on the walkers and weighted by some positive function $\psi_G$
\begin{equation}
 f_t ({\bf R}) =  \frac{1}{\psi_G ({\bf R})} \left\langle \frac{}{} \sum_i
 \delta ({\bf R}-{\bf R}_i ) \right\rangle
\label{sampleft}
\end{equation}
In this expression, $\langle ... \rangle$ denotes the average over populations of 
walkers $\{{\bf R}_i\}$ obtained at the given time $t$. 
The function $\psi_G$ is usually called the importance or guiding function. 
Replacing $f_t$ in (\ref{E_Fproj}) by its expression (\ref{sampleft}), 
the estimator of the energy reads for large enough $t$
\begin{eqnarray}
E_0^F & = & 
\frac{\left\langle \sum_i  \frac{H\psi_T}{\psi_G} ({\bf R}_i) \right\rangle}
     {\left\langle \sum_i \frac{\psi_T}{\psi_G} ({\bf R}_i)\right\rangle}
\label{energystat}
\end{eqnarray}
In practice, both numerator and denominator are computed as an average
over the  $N_S$ walkers produced by the algorithm at time $t$.
As a consequence, the energy is obtained as
\begin{equation}
E_0^F = \frac{\cal{N} }{\cal{D}}
\equiv  \frac{ \frac{1}{N_S}\sum_{i=1}^{N_S}   \frac{H\psi_T}{\psi_G} ({\bf
    R}_i)}{ \frac{1}{N_S} \sum_{i=1}^{N_S}   \frac{\psi_T}{\psi_G} ({\bf R}_i)}.
\label{statea}
\end{equation}
The ratio $\frac{\cal{N} }{\cal{D}}$ is an estimator of the energy for $N_S$ large enough, 
when the numerator and denominator have small fluctuations aroung their average.
Now, let us evaluate the fluctuations of this ratio in the limit of a large population $N_S$
\begin{equation}
\sigma^2 \left( \frac{\cal{N}}{\cal{D}} \right) = \frac{\langle
  ({\cal N}-E_0^F{\cal D})^2 \rangle}{\langle {\cal D} \rangle^2}.
\end{equation}
Here, we have used the hypothesis that the fluctuations of the denominator and the 
numerator are very small.
Using the fact that ${\cal N}$ and ${\cal D}$ are statistical averages 
over independent random variables 
with the same distribution one obtains
 \begin{equation}
\sigma^2 \left( \frac{\cal{N}}{\cal{D}} \right) = \frac{1}{N_S} \frac{\langle  
\left[\frac{(H-E_0^F)\psi_T}{\psi_G} ({\bf R}_i)\right]^2  \rangle}{ \langle  \frac{\psi_T}{\psi_G} ({\bf R}_i) \rangle^2}.
\end{equation}

We can replace these averages by integrals over the  distribution  $f_t \psi_G$, 
Eq.(\ref{sampleft})
\begin{equation}
\sigma^2 \left( \frac{\cal{N}}{\cal{D}} \right) = \frac{1}{N_S} 
\frac{\langle  \frac{[(H-E_0^F)\psi_T]^2}{\psi_G} \mid \phi_0^B  \rangle
\langle \psi_G \mid \phi_0^B  \rangle }{ \langle \psi_T \mid \phi_0^F  \rangle^2}
e^{2t(E_0^F-E_0^B)} \propto \frac{1}{N_S} e^{2t(E_0^F-E_0^B)}
\label{sigmaea}
\end{equation} 
which confirms quantitatively that the statistical error grows exponentially
in time. Let us emphasize that this problem is particularly severe because the Bose-Fermi 
energy gap, $\Delta_{B-F} \equiv E_0^F-E_0^B$, 
usually grows faster than linearly as a function of the number of particles.
In practice, one has to find a trade-off between the systematic error coming from 
short projection times $t$ and the large fluctuations arising at large projection times.

\section{The diffusion Monte Carlo method}
The FMC method is a generalization of the well-known  DMC
method. Presenting this algorithm is a useful preparation
for the next section about Fermion Monte Carlo.
The DMC method  generates the function $f_t$ following the imaginary time dymamics of $H$
\begin{equation}
f_t \equiv e^{-t(H-E_T)} f_0
\label{evolt0}
\end{equation}
where $f_0$ is  a positive function.
The imaginary time dynamics is produced by iterating many times the short-time
Green function  $e^{-\tau (H-E_T)}$ where $\tau$ is a small time step.
The distribution $f_{t^\prime}$ at the time $t^\prime \equiv t+\tau$ is then obtained from $f_t$ as follows 
\begin{equation}
f_{t^\prime} = e^{-\tau (H-E_T)} f_t
\label{shorttime}
\end{equation}
where the density $f_t$ is sampled by the population of walkers $\{ {\bf R}_i\}$.
Using the Dirac ket notation, $f_t$ given  by Eq.(\ref{sampleft}) is rewritten as
\begin{equation}
f_t  =  \frac{1}{\psi_G} \left\langle \frac{}{} \sum_i \mid {\bf R}_i \rangle \right\rangle
\label{def_ft2},
\end{equation}
where $\psi_G$ is some positive function, the so-called guiding function.
Let us show how the density $f_{t^\prime}$ is generated from the distribution (\ref{def_ft2}).
Replacing in (\ref{shorttime}) the function $f_t$ by the R.H.S. of (\ref{def_ft2}) one has 
\begin{eqnarray}
 f_{t^\prime} & = &  e^{-\tau (H-E_T)} \frac{1}{\psi_G}
 \left\langle  \frac{}{} \sum_i  \mid {\bf R}_i \rangle \right\rangle \\
  & = &
  \left\langle \sum_i  e^{-\tau (H-E_T)} \frac{1}{\psi_G} \mid
  {\bf R}_i \rangle \right\rangle \\
  & = &  \frac{1}{\psi_G} \left\langle  \frac{}{} \sum_i e^{\tau L} \mid {\bf  R}_i \rangle \right\rangle
\label{dmcopap}
  \end{eqnarray}
where we have introduced the operator
\begin{eqnarray}
  L & \equiv &  -\psi_G (H-E_T) \frac{1}{\psi_G}.
 \label{dmcop}
\end{eqnarray}
For a Schr\"odinger Hamiltonian (\ref{schro}) the operator $L$ takes the form
\begin{eqnarray}
L & = & -\psi_G (H-E_L) \frac{1}{\psi_G} - (E_L-E_T) \\
  & = &
\frac{1}{2} {\bf \nabla}^2 - {\bf \nabla} [{\bf b}.] 
\label{fkp} \\
  &   & - (E_L -E_T)
\label{branching}
\end{eqnarray}
where   we have introduced the so-called drift vector
\begin{equation}
{\bf b} \equiv \frac { {\bf \nabla} \psi_G}{\psi_G}
\end{equation}
and the local energy of the guiding function $\psi_G$, 
\begin{equation}
E_L \equiv \frac{H \psi_G}{\psi_G}.
\end{equation}
The operator $L$ is the sum of the so-called Fokker Planck operator  (\ref{fkp}) and a local 
operator (\ref{branching}). Using this decomposition, 
the vector $e^{\tau L} \mid {\bf R}_i \rangle$ appearing in the 
average (\ref{dmcopap}) can be rewritten for small enough time step $\tau$ as follows
\begin{eqnarray}
e^{\tau L} \mid {\bf R}_i \rangle & = &
e^{\tau \left[  \frac{1}{2} {\bf \nabla}^2 - {\bf \nabla} [{\bf b}.] \right]} 
\label{fkp_ap}
\\
& & \times e^{-\tau (E_L-E_T)} \mid {\bf R}_i \rangle.
\label{branch_ap}
\end{eqnarray}
The action of $e^{\tau L}$ on $ \mid {\bf R}_i \rangle $ is sampled as follows. 
The short-time dynamics of the Fokker Planck operator (\ref{fkp_ap}) is performed by the way of 
 a Langevin process,
\begin{equation}
{{R_i}^\prime}^\mu = {R_i}^\mu + b_i^\mu  \tau + \sqrt{\tau} \eta_i^\mu
\label{lang}
\end{equation}
where $\mu$ runs over the $3N$ coordinates (three space coordinate for each 
fermion), and $\eta_i^\mu$ are independent gaussian random variables centered and normalized
 \begin{equation}
\langle \eta_i^\mu \eta_i^\nu \rangle =\delta^{\mu\nu} 
 \end{equation}
The averaged Langevin process (\ref{lang}) is equivalent to apply 
the short-time dynamics of the Fokker Planck operator (\ref{fkp_ap}):
\begin{equation}
\left\langle  \frac{}{} \mid {\bf R}_i^\prime \rangle \right\rangle =e^{\tau \left[  \frac{1}{2} 
{\bf \nabla}^2 - {\bf \nabla} [{\bf b}.] \right]}  \mid {\bf R}_i \rangle
\end{equation}
 The  factor 
\begin{equation}
w_i \equiv e^{-\tau (E_L ({\bf R_i})-E_T)}
\label{branching2}
\end{equation}
 being a normalization term, called the
 branching term. The new walker ${\bf R}_i^\prime$ is duplicated (branched)
 a number of times equal to $w_i$ in average.
This process is a birth-death process since some walkers can be duplicated
 and some can be removed. The population of walkers
 fluctuating, one has to resort to control
 population techniques. \cite{umrigarJCP93,sorella98rec,khelif2000}
With these two processes, diffusion and branching 
a new population of walkers $\{{\bf R}_i^\prime\}$ is produced, representing in
 average the desired result :
\begin{equation}
\left\langle \frac{}{} \sum_i \mid {\bf R}_i^\prime \rangle \right\rangle
= \left\langle  \sum_i e^{\tau L} \mid {\bf R}_i \rangle \right\rangle
\label{riprim}
\end{equation}
From (\ref{dmcopap}) and (\ref{riprim}) one can see that 
the distribution $f_{t^\prime}$ is sampled by the new population of 
walkers $\{{\bf R}_i^\prime\}$ according to (\ref{def_ft2})
\begin{equation}
f_{t^\prime} =   \left\langle  \frac{1}{\psi_G} \mid {\bf R}_i^\prime) \right\rangle.
\label{ft_samplep}
\end{equation}
In summary, by iterating these two simple operations, namely the Langevin and 
branching processes, the DMC method allows to simulate the imaginary time dynamics 
of the Hamiltonian, thus producing a sample of $f_t$, Eq.(\ref{def_ft2}).
Various properties of the system can be computed from this sample, e.g. 
ground-state bosonic energies,\cite{klv,ceperleykalos79,schmittkalos84}
excited-state energies,\cite{bernuJCP91,caffarelJCP1992} and various observables.
\cite{rothstein97,Rothstein2004,force00,filippi2000,caffrerat,rcaf03,casalegno2003}

For the vast majority of the DMC simulations on fermionic systems, only an 
approximation of the exact 
fermionic ground-state energy is computed, namely the so-called Fixed-Node energy.
\cite{schmittkalos84,reynoldsJCP82,filippiJCP96,ceperleyalder86,
ceperleyalderPRL80,Grossman2002} 
In a fixed-node DMC calculation the guiding function is chosen as
$\psi_G=|\psi_T|$ where $\psi_T$ is some fermionic antisymmetric trial wavefunction.
With this choice, the guiding function 
vanishes at the nodes (zeroes) of the trial wavefunction and the drift vector diverges.
As a consequence, the walkers cannot cross the nodes of $\psi_T$
and are confined within the nodal regions of the configuration space.
It can be shown that the resulting DMC stationary state is the best variational
solution having the same nodes as $\psi_T$. In other words, the ``Fixed-Node'' energy obtained 
from the R.H.S of (\ref{energystat}) or (\ref{statea})  
is an upper bound of the exact fermi energy, 
$E_0^{FN}>E_0^F$.\cite{ceperleyalderPRL80,ceperleyalder86}
Note that, in practice, $E_0^{FN}$ is in general a good approximation of 
the true energy. \cite{filippiJCP96,Grossman2002}

In the present work, we are considering ``exact'' DMC approaches for which the exact 
fermionic energy calculated from expression (\ref{energystat}) is searched for.
As discussed in the previous section, such exact DMC calculations are fundamentally unstable.
A famous example of an exact DMC approach is the 
nodal release method of Ceperley and Alder.\cite{ceperleyalderPRL80,ceperleyalderPRL84}
Basically, nodal release methods are standard DMC methods where 
the fixed-node distribution (sampled with a standard Fixed-Node
DMC) is chosen as initial distribution $f_0$.
In exact methods the guiding function $\psi_G$ is strictly positive
everywhere, so that the walkers can cross the nodes of the trial wavefunction.
Exact fermi methods are efficient in practice only when
the convergence of the estimator is fast enough, that is, when it occurs before 
the blowing up of fluctuations, Eq. (\ref{sigmaea}). 
In practice, two conditions are to be satisfied.
First, the fixed-node wavefunction
$\phi_0^{FN}$ must be already close enough to the exact solution $\phi_0^F$.  For this
reason the choice of the trial function $\psi_T$ (quality of the nodes of $\psi_T$)
is crucial.
Second, the Bose-Fermi gap, $\Delta_{B-F} = E_0^F-E_0^B$, 
which drives the asymptotic behaviour of the
fluctuations, Eq. (\ref{sigmaea}), must be small.
The quantity $E_0^F-E_0^B$ depends only on the Hamiltonian at hand;  there is
no freedom in the nodal release method to modify the asymptotic behaviour.
We will define in the next section the Fermion Monte Carlo method (FMC) as
a generalization of the DMC method and show in sections V and VI
that, in contrast with the nodal release method, the FMC method can improve 
substantially the asymptotic behaviour, Eq.(\ref{sigmaea}).

\section{The FMC method}
\subsection{ Preliminary: Introducing positive and negative walkers in DMC}
In Fermion Monte Carlo a dynamics on a signed function $f_t$ is performed.
In what follows, we show that DMC can be easily generalized to the case of 
a signed distribution $f_t$ and, thus, FMC  can be viewed as 
%This formulation has  usually no interest if we are interested in  the
%asymptotic behaviour  of $f_t$, which is a positive function,  
%namely the lowest eigenstate  of $H$.
%Here this formulation may have an interest because we are looking at the transient behaviour.
a simple generalization of DMC.
%End of RA
If $f_t$ carries a sign, it can be written as the difference of two positive functions
\begin{equation}
f_t = f_t^+-f_t^-
\label{f_tdif}
\end{equation}
both satisfying the following equations of evolution
\begin{eqnarray}
f_t^+ & = & e^{-t(H-E_T)} f_0^+ \\
f_t^- & = &  e^{-t(H-E_T)} f_0^-
\end{eqnarray}
To sample these expressions, two independent DMC calculations 
can be carried out. The positive part $f_t^+$ is then sampled by a population of 
walkers $\{{\bf   R_i}^+\}$ (called ``positive'' walkers). The distribution $f_t^+$ 
is related to  $\{{\bf R_i}^+\}$ as in Eq.(\ref{def_ft2})
\begin{equation}
f_t^+   =   \frac{1}{\psi_G^+} \left\langle \frac{}{} \sum_i  \mid {\bf R}_i^+ \rangle \right\rangle
\label{sampleft+}
\end{equation}
and the negative part is similarly sampled by a population of 
``negative'' walkers $\{{\bf R_i}^-\}$
\begin{equation}
f_t^-   =  \frac{1}{\psi_G^-} \left\langle \frac{}{} \sum_i \mid {\bf R}_i^- \rangle \right\rangle.
\label{sampleft-}
\end{equation}
Note that we consider here the general case 
where the guiding functions associated with the positive and negative walkers, 
$\psi_G^+$ and  $\psi_G^-$, are different.
Finally, the dynamics of positive and negative walkers is described by the DMC-like diffusion 
operators
\begin{eqnarray}
  L^\pm & \equiv &  -\psi_G^{\pm} (H-E_T) \frac{1}{\psi_G^{\pm}} 
  \\
  & = &  \frac{1}{2} {\bf \nabla}^2 - \nabla [{\bf b^\pm}.] 
\label{fkpp} \\
  &   & - (E_L^\pm -E_T) \label{branchingp}
\end{eqnarray}
where  the  drift vectors are given by
\begin{equation}
{\bf b}^\pm \equiv \frac { {\bf \nabla} \psi_G^\pm}{\psi_G^\pm}
\end{equation}
and the local energies of the guiding functions $\psi_G^\pm$ by
\begin{equation}
E_L^\pm \equiv \frac{H \psi_G^\pm}{\psi_G^\pm}.
\label{elgw}
\end{equation}
In actual calculations, two Langevin dynamics on the positive and negative walkers are 
performed
\begin{equation}
{{{R_i}^\prime}^\mu}^\pm = {{R_i}^\mu}^\pm + b^\mu \tau + \sqrt{\tau} {\eta_i^\mu}^\pm
\label{lang+-}
\end{equation}
and positive and negative walkers are branched according to  their respective weight
\begin{equation}
W^\pm ({\bf R}_i^\pm) \equiv e^{-\tau ({E_L}^\pm ({\bf R}^\pm_i)-E_T)}.
\label{FKW+-}
\end{equation}
\subsection{The detailed rules of FMC}
In a few words, the FMC method is similar to a DMC method on a 
signed function, except that the positive and negative walkers are correlated and can 
annihilate whenever they meet. The Langevin processes are correlated in such a way that 
positive walkers and negative walkers meet as much as possible.
A cancellation procedure is then performed when a positive walker and a negative 
walker meet. We will see in the next section that this cancellation procedure is 
at the origin of an improved stability of the algorithm. In this section we give 
a complete description of the algorithm and show that this algorithm does not 
introduce any systematic bias.

In FMC the guiding functions $\psi_G^+$ and $\psi_G^-$ are not arbitrary,
they are related under any permutation $P$ of two particles as follows
\begin{equation}
\psi_G^+ ({\bf R}) = \psi_G^- ( P{\bf R}).
\label{ppsig+-}
\end{equation}
Various choices are possible for the guiding functions. Here, we consider the 
form proposed by Kalos and Pederiva,\cite{kalosfermion00} namely
\begin{equation}
\psi_G^{\pm} = \sqrt{\psi_S^2 + c^2 \psi_T^2}\pm c \psi_T
\label{psig+-def}
\end{equation}
where $\psi_S$ is a symmetric (Bose-like) function, $\psi_T$ an antisymmetric trial wavefunction, 
and $c$ some positive mixing parameter allowing to introduce some antisymmetric component into 
$\psi_G^{\pm}$.

Having in mind this choice for the guiding functions we will show how the two
DMC processes over the two populations of walkers  $\{{\bf R}_i^+\}$ and $\{{\bf
R}_i^-\}$ can be replaced by a diffusion process over a population of 
{\it pairs} of walkers $\{({\bf R}_i^+,{\bf R}_i^-)\}$. 
We first show how the DMC process can be modified to maintain as many
positive and negative walkers during the simulation.
In the DMC dynamics, the branching terms associated with the positive and negative walkers 
are in general different. 
As a consequence, the number of positive walkers  $N_S^+$, can be different from the 
number of negative walkers $N_S^-$. At time $t$ the DMC density $f_t$ reads 
\begin{equation}
f_t = \left\langle \sum_{i=1}^{N_S^+}  \frac{1}{\psi_G^+} \mid {\bf R}_i^+ \rangle -  \sum_{i=1}^{N_S^-} \frac{1}{\psi_G^-} \mid {\bf R}_i^- \rangle \right\rangle.
\label{dmcsample+-}
\end{equation}
This formula is obtained by replacing in equation (\ref{f_tdif}) the expressions 
(\ref{sampleft+}) and (\ref{sampleft-}) for
$f_t^+$ and $f_t^-$, respectively.
If $N_S^+$ and $N_S^-$ are different, 
we will replace $f_t$ (\ref{dmcsample+-}) by a 
new function $g_t$ sampled with an equal number of positive and negative walkers.
Such an operation does not introduce any bias if the antisymmetric 
components of the future evolution of $f_t$ and $g_t$ are identical. 
Indeed, only the antisymmetric component of $f_t$ contributes to the 
estimator of the energy, Eq. (\ref{E_Fproj}).
At time $t^\prime > t$ the two densities are given by
\begin{eqnarray}
 f_{t^\prime}  & = & e^{-(t^\prime -t) (H-E_T)} f_t \\
 g_{t^\prime} & =  &  e^{-(t^\prime -t) (H-E_T)} g_t.
\end{eqnarray}
Let us write that the  antisymmetric components of $f_{t^\prime}$ and $g_{t^\prime}$ must be equal
\begin{equation}
{\cal A} e^{-(t^\prime -t) (H-E_T)} f_t  = {\cal A} e^{-(t^\prime -t) (H-E_T)} g_t.
\end{equation}
Using the fact that the antisymmetrisation operator ${\cal A}$ commutes with the evolution
operator and regrouping all the terms one finally finds 
\begin{equation}
 e^{-(t^\prime -t) (H-E_T)} {\cal A} (f_t-g_t)  = 0.
\end{equation}
This condition is satisfied whenever ${\cal A} (f_t-g_t)  = 0$. 
The important conclusion is that one can replace $f_t$ by any function $g_t$ 
such that the difference $g_t-f_t$ is 
orthogonal to the space of antisymmetric functions.
Let us now show how this property can be used to impose a common number of walkers in 
the positive and negative populations.

Let us consider the case where there are more positive walkers than negative walkers,
$N_S^+ > N_S^-$.  
In this case one can substract from $f_t$, Eq.(\ref{dmcsample+-}), the following vector
\begin{equation}
\frac{1}{\psi_G^+} \mid {\bf R}_i^+ \rangle + P  \frac{1}{\psi_G^+} \mid {\bf R}_i^+ \rangle 
\label{1+P+}
\end{equation}
where ${\bf R}_i^+$ is a positive walker and $P$ is a two-particle permutation.
Such an operation is allowed since the application of the antisymmetrizer to the 
vector (\ref{1+P+}) gives zero [a direct consequence of ${\cal A}(1 + P)=0$].
Now, because of Eq.(\ref{ppsig+-}) the vector (\ref{1+P+}) can also be written as
\begin{equation}
\frac{1}{\psi_G^+} \mid {\bf R}_i^+ \rangle +   \frac{1}{\psi_G^-} P \mid {\bf R}_i^+ \rangle.
\end{equation}
Substracting this vector from $f_t$ removes the contribution $\frac{1}{\psi_G^+} \mid {\bf R}_i^+ \rangle$ from (\ref{dmcsample+-}) and adds the contribution $- \frac{1}{\psi_G^-} P \mid {\bf R}_i^+ \rangle$ to (\ref{dmcsample+-}).
In other words, the positive walker ${\bf R}_i^+$ has been removed and the negative walker
\begin{equation}
{\bf R}_i^- = P {\bf R}_i^+
\end{equation}
has been created.
Similarly, one can remove a negative walker ${\bf R}_i^-$ and create a positive walker 
\begin{equation}
{\bf R}_i^+ = P {\bf R}_i^-.
\end{equation}
This possibility of transfering one walker from one population to the other one allows to 
keep an identical number of walkers for the two populations at each step. 
Now, thanks to this possibility, it is possible to
interpret the two populations consisting of
the $N_S$ positive walkers ${\bf R}_i^+$ and  the $N_S$ negative walkers 
${\bf R}_i^- $ ($i \in [1..N_S]$) 
as an unique population of $N_S$ pairs of walkers $\{({\bf R}_i^+,{\bf R}_i^-)\}$. 
Following this interpretation, the density $f_t$ (\ref{dmcsample+-}) can be then rewritten 
as an average over a population of pairs of walkers
\begin{equation}
f_t = \left\langle \sum_{i=1}^{N_S}  (\frac{1}{\psi_G^+} \mid {\bf R}_i^+ \rangle 
-  \frac{1}{\psi_G^-} \mid {\bf R}_i^- )\rangle \right\rangle
\label{f_tfmc}
\end{equation}
and the energy can be computed as a ratio of averages performed on the population of pairs 
\begin{equation}
E_0^F = \frac{\left\langle \sum_i (\frac{H\psi_T}{\psi_G^+} ({\bf R}_i^+) - 
\frac{H\psi_T}{\psi_G^-} ({\bf R}_i^-)) \right\rangle}
     {\left\langle \sum_i (\frac{\psi_T}{\psi_G^+} ({\bf R}_i^+)- 
\frac{\psi_T}{\psi_G^-} ({\bf R}_i^-))\right\rangle},
\label{eapair}
\end{equation}
where Eq.(\ref{E_Fproj}) has been rewritten by replacing $f_t$ using Eq.(\ref{f_tfmc}).
Now, everything is in order to detail the short-time dynamics of FMC.
The FMC dynamics consists of three steps (Langevin, branching, and 
cancellation steps):

(i) {\sl  Langevin step}
The Langevin processes (\ref{lang+-}) are simulated as in DMC, except that the gaussian 
random variables of the positive and negative walkers are no longer independent.
The positive walker ${\bf R}_i^+$ and the negative walker ${\bf R}_i^-$ are moved according 
to Eq.(\ref{lang+-})
\begin{equation}
{{{R_i}^\prime}^\mu}^\pm = {{R_i}^\mu}^\pm + {b^\mu}^\pm \tau + \sqrt{\tau} {\eta_i^\mu}^\pm
\label{lang+-2}
\end{equation}
where ${\eta_i^\mu}^\pm$ are gaussian centered random variables verifying
\begin{equation}
\langle {\eta_i^\mu}^\pm {\eta_j^\mu}^\pm \rangle = \delta_{ij}.
\end{equation}
Such a move insures that the density of positive and negative walkers obey the Fokker Planck 
equation: 
\begin{equation}
\left\langle \frac{}{} \mid {{\bf R}_i^\prime}^\pm \rangle \right\rangle = e^{\tau ( \frac{1}{2}
{\bf \nabla}^2
- {\bf \nabla} [{\bf b^\pm}.])} \mid {\bf R}_i^\pm  \rangle.
\end{equation}

However, the gaussian random variables are no more independent, 
they are correlated within a pair 
\begin{equation}
c_i^{\mu \nu} \equiv \langle {\eta_i^\mu}^+ {\eta_i^\nu}^- \rangle \ne 0.
\end{equation}
Differents ways of correlating positive and negative walkers can be considered.
We shall employ here the approach used in Refs.\cite{Liu1994,kalosfermion00}
which consists in obtaining 
the vector $\vec{{\bf \eta}_i}^-$, representing the $3N$ 
coordinates ${\vec{\bf \eta}_i}^-$, from ${\vec{\bf \eta}_i}^+$ by reflexion 
with respect to the hyperplane perpendicular to the vector ${\bf R}_i^+-{\bf R}_i^-$.
\begin{equation}
\vec{\eta_i} ^- = \vec{\eta_i}^+ - 2 \frac{({\bf R}_i^+-{\bf R}_i^-).\vec{\eta_i} ^+}{({\bf R}_i^+-{\bf R}_i^-)^2} ({\bf R}_i^+-{\bf R}_i^-).
\label{cor+-}
\end{equation}
This relation between the gaussian random variables makes the move deterministic 
along the direction ${\bf R}_i^+-{\bf R}_i^-$. Such a construction insures that the walkers 
within a pair will meet each other in a finite time (even in large-dimensional spaces). 
This aspect will be illustrated numerically in the last section.
Formally, the two correlated Langevin processes can be seen as one Langevin process 
in the space of pairs of walkers
\begin{equation}
({{\bf R}_i^+}^\prime,{{\bf R}_i^-}^\prime) =  ({{\bf R}_i^+},{{\bf R}_i^-}) + ({\bf b^+}({\bf R}_i^+), {\bf b^-({\bf R}_i^-)})\tau + \sqrt{\tau} (\vec{\eta}_i^+,\vec{\eta}_i^-)
\label{corrlanffmc}
\end{equation}
where $\vec{\eta}_i^+$ and $\vec{\eta}_i^-$ are related via (\ref{cor+-}).

{\sl (ii) Branching step}
As we have already noticed, the branching of the negative and the positive 
walker are different within a pair:
\begin{equation}
w_i^+ \equiv e^{-\tau (E_L^+ ({\bf R}_i^+)-E_T)} \ne w_i^-   \equiv e^{-\tau (E_L^- ({\bf R}_i^-)-E_T)}.
\end{equation}
Taking into account their respective weights, the two walkers of a pair give the 
following contribution to the density $f_t$ 
\begin{equation}
w_i^+ \frac{1}{\psi_G^+} \mid {\bf R}_i^+ \rangle - 
w_i^-  \frac{1}{\psi_G^-} \mid {\bf R}_i^- \rangle .
\label{contw+-}
\end{equation}
If for example $w_i^+>w_i^-$, this vector can be written as
\begin{eqnarray}
& & w_i^- \left[\frac{1}{\psi_G^+} \mid {\bf R}_i^+ \rangle -\frac{1}{\psi_G^-} \mid {\bf R}_i^- \rangle \right] 
\label{branchingpair+-} \\
& & + (w_i^+-w_i^-) \frac{1}{\psi_G^+} \mid {\bf R}_i^+ \rangle
\label{branchingsingle}
\end{eqnarray}
This density is the sum of two contributions. The first contribution, Eq.(\ref{branchingpair+-}), 
comes from a pair of walkers $({\bf R}_i^+, {\bf R}_i^-)$  and carries the weight $w_i^-$.
The second, Eq.(\ref{branchingsingle}), comes from a single 
positive walker ${\bf R}_i^+$  and carries the weight $w_i^+-w_i^-$. 
This single walker  ${\bf R}_i^+$ can be replaced by a pair as follows.
First, this single
 walker can be replaced by two positive walkers ${\bf R}_i^+$ with half of the
 weight, $\frac{1}{2}(w_i^+-w_i^-)$. One of these two positive walkers carrying
 half of the weight can be transfered to the population of negative walkers by
 exchanging two particles.
Finally, this single walker ${\bf R}_i^+$ can be replaced by a pair 
$({\bf R}^+_i,P{\bf R}^+_i)$  carrying the weight $\frac{1}{2}(w_i^+-w_i^-)$.
The resulting process just described is a branching of the pair 
$({\bf R}_i^+, {\bf R}_i^-)$ with the weight $w_i^-$ and the creation of 
the pair $({\bf R}^+_i,P{\bf R}^+_i)$  with the weight $\frac{1}{2}(w_i^+-w_i^-)$.
Of course, if one has $w_i^+<w_i^-$, then, the pair $({\bf R}_i^+, {\bf R}_i^-)$  is 
branched with the weight $w_i^+$ and the pair $(P{\bf R}^-_i,{\bf R}^-_i)$ is created 
with the weight $\frac{1}{2}(w_i^--w_i^+)$.

Both cases, $w_i^-<w_i^+$ or $w_i^->w_i^+$,  can be summarized as follows.
The pair of walkers $({\bf R}_i^+,{\bf R}_i^-)$ is branched  with the weight
\begin{equation}
\text{min} (w_i^+,w_i^-) = e^{-\tau (\text{max} (E_L^+,E_L^-)-E_T)}
\label{wminpair}
\end{equation}
and the pairs $({\bf R}^+_i,P{\bf R}^+_i)$ and $(P{\bf R}^-_i,{\bf R}^-_i)$
are created with their respective weights
\begin{equation}
\frac{1}{2}(w_i^+-\text{min}(w_i^+,w_i^-)) = \frac{\tau}{2} [E_L^+-\text{max}
  (E_L^+,E_L^-)]
+ O(\tau^2)
\label{wsingle1}
\end{equation}
and
\begin{equation}
\frac{1}{2}(w_i^--\text{min}(w_i^+,w_i^-)) = \frac{\tau}{2} [E_L^+-\text{max}
  (E_L^+,E_L^-)]
+O(\tau^2)
\label{wsingle2}
\end{equation}

{\sl (iii) Cancellation step}
The third step is a cancellation procedure performed whenever a positive and negative walker meet.
When ${\bf R_i}^+={\bf R_i}^-$, the contribution of the pair to the density can be simplified 
as follows
\begin{equation}
\left[1-\frac{\psi_G^+}{\psi_G^-}({\bf R}_i^+)\right] \frac{1}{\psi_G^+}({\bf R}_i^+) \mid {\bf R}_i^+ \rangle.
\end{equation}
If the term in brackets is positive, this contribution comes from one single positive 
walker ${\bf R}_i^+$ with multiplicity $\left[1-\frac{\psi_G^+}{\psi_G^-}({\bf R}_i^+)\right]$.
One can transform this single walker into a pair of positive and negative walker 
$({\bf R}_i^+,P{\bf R}_i^+)$ with the new multiplicity
\begin{equation}
\frac{1}{2}\left[1-\frac{\psi_G^+}{\psi_G^-}({\bf R}_i^+)\right].
\label{multc+}
\end{equation}
If the term in brackets is negative,  the pair  $(P{\bf R}_i^+,{\bf R}_i^+)$ is drawn and 
the multiplicity is given by
\begin{equation}
\frac{1}{2}\left[1-\frac{\psi_G^-}{\psi_G^+}({\bf R}_i^+)\right].
\label{multc-}
\end{equation}
This is a cancellation procedure because a pair 
$({\bf R}_i^+,{\bf R}_i^+)$ with a multiplicity 1 has been transformed into a pair 
with a multiplicity smaller than one.
Note that, when $\psi_G^+=\psi_G^-$, the multiplicities (\ref{multc+}) or 
(\ref{multc-}) reduce both to zero. In other words, there is a total cancellation 
of the pair whenever the walkers meet.
As we shall see in the next section, the cancellation step is 
at the origin of the improved stability. The basic reason is that this
procedure removes pairs which do not contribute to the signal but only to the statistical noise.
A rigorous analysis of this point is provided in the next section.

\section{Stability of the FMC method}
\subsection{Criterium for stability}
We have just seen that the FMC method is a generalization of the DMC approach and
we have shown that FMC preserves the evolution of the antisymmetric component of the
sampled density.
Now, having shown that FMC is an exact method, it is necessary to study the stability 
of the method.
For that purpose, we consider the estimator of the energy, Eq. (\ref{eapair}), in the large-time 
regime
\begin{equation}
E_0^F = \frac{\cal N}{\cal D} = 
\frac{\frac{1}{N_S} \sum_i  \frac{H\psi_T}{\psi_G^+} ({\bf R}_i^+) - \frac{H\psi_T}{\psi_G^-} ({\bf R}_i^-) }
     { \frac{1}{N_S} \sum_i \frac{\psi_T}{\psi_G^+} ({\bf R}_i^+)- \frac{\psi_T}{\psi_G^-} ({\bf R}_i^-)}
\label{efmcN/D}
\end{equation}
where $N_S$ is the population size at time $t$.
In the same way as done
for the estimator (\ref{statea}), we can evaluate the variance of $E_0^F$ by supposing 
that $N_S$ is large enough
so that both numerator and denominator have small fluctuations around their average
\begin{equation}
\sigma^2 \left( \frac{\cal{N}}{\cal{D}} \right) = \frac{\langle
  ({\cal N}-E_0^F{\cal D})^2 \rangle}{\langle {\cal D} \rangle^2}.
\label{sigmaefmc}
\end{equation}
Let us begin with the denominator.
Using identity (\ref{f_tfmc}), the average of the random variable ${\cal
 D}$ defined in(\ref{efmcN/D}) is nothing but 
\begin{equation}
\langle {\cal D} \rangle = \frac{1}{N_S} \langle \psi_T \mid f_t \rangle.
\end{equation}
Replacing $f_t$ by its asymptotic behaviour, Eq.(\ref{aft}), we
finally find that the denominator of (\ref{sigmaefmc}) behaves for large $t$ as 
\begin{equation}
\langle {\cal D} \rangle^2 =  \frac{1}{N_S^2} e^{-2t(E_0^F-E_T)} \langle \psi_T
\mid \phi_0^F \rangle^2.
\label{denomfmcbehaviour}
\end{equation}
Now, let us compute the numerator of (\ref{sigmaefmc}). This numerator can be
written as the variance of a sum of random variables defined over the pairs of walkers
\begin{equation}
\langle  ({\cal N}-E_0^F{\cal D})^2 \rangle =
\left\langle \left[ \frac{ \sum_i \Gamma({\bf R^+}_i,{\bf R}^-_i)}{N_S}  \right]^2 \right\rangle
\end{equation}
where we have introduced the function $\Gamma({\bf R}^+,{\bf R}^-)$
\begin{equation}
\Gamma({\bf R}^+,{\bf R}^-)\equiv \frac{(H-E_0^F)\psi_T}{\psi_G^+} ({\bf R}^+) -
  \frac{(H-E_0^F)\psi_T}{\psi_G^-} ({\bf R}^-).
\end{equation}
Using the fact that the pairs of walkers have the same distribution and supposing that
they are independent we finally find
\begin{equation}
\langle  ({\cal N}-E_0^F{\cal D})^2 \rangle =  \frac{1}{N_S}  \sigma^2 \left(\frac{}{}\Gamma({\bf R}^+,{\bf R}^-)\right).
\end{equation}
If the pairs of walkers are not independent
this expression is only modified by a correlation factor 
independent on the time and the population size $N_S$
provided that $N_S$ and $t$ are large enough.
Finally, up to a multiplicative factor, the variance
of the energy estimator has the following asymptotic behaviour 
\begin{equation}
\sigma^2(\frac{\cal N}{\cal D})  \propto  \frac{1}{N_S(t)} C(t)
\label{sigma2efmc}
\end{equation}
where the coefficient $C(t)$ is given by
\begin{equation}
C(t) = N_S(t)^2 e^{-2t(E_0^F-E_T)}
\label{C(t)}
 \end{equation}
and where $N_S(t)$, the number of pairs, depends on time $t$ due to the
birth-death process.
In expression (\ref{C(t)}) we note that the behaviour of (\ref{sigmaefmc}) at
large times $t$ is related to the value of $E_0^F$ and the asymptotic behaviour of
the number of pairs of walkers $N_S(t)$.
We can already understand physically the interest of
the cancellation process: this process limits the growth of $N_S(t)$ the number of
pairs of walkers, thus limiting the growth of $C(t)$ (\ref{C(t)}) and the
variance (\ref{sigma2efmc}).
Let us now precise this criterium more rigorously  by evaluating the
asymptotic behaviour of $N_S(t)$.
For that purpose we introduce the density of pairs $\Pi_t ({\bf R}_i^+,{\bf R}_i^-)$, this
density obeys a diffusion equation we write down
\begin{equation}
\frac{\partial \Pi_t}{\partial t} = - ({ D}_{\text{FMC}}-E_T) \Pi_t
\label{evotilde}
\end{equation}
where we have introduced the diffusion operator $-({ D}_{\text{FMC}}-E_T)$.
We will give its expression later; for the present purpose we just need
the asymptotic behaviour of $\Pi_t$ given by
\begin{equation}
\Pi_t = e^{-t(\tilde{E}_0^B-E_T)} \Pi_S
\label{asympit}
\end{equation}
where $\Pi_S$ is the stationary density of the process, namely the
lowest eigenstate of the operator $D_{\text{FMC}}$, and $\tilde{E}_0^B$ is the
corresponding eigenvalue.
The number of pairs $N_S(t)$ behaves as the normalization of $\Pi_t$, and consequently grows like
$e^{-t(\tilde{E}_0^B-E_T)}$.
Note that in practice one adjusts the reference energy $E_T$ to $\tilde{E}_0^B$ during
the simulation to keep a constant population of average size $\bar{N_S}$ along the dynamics.
Such a procedure is referred to as a control population technique\cite{khelif2000} 
and will be discussed later.
Finally, the asymptotic behaviour of the variance of the FMC estimator of the energy is
\begin{equation}
\sigma^2(\frac{\cal N}{\cal D}) \propto \frac{1}{\bar{N_S}}  e^{2t(E_0^F-\tilde{E}_0^B)}
\label{fluctfmc}
\end{equation}
This expression is analogous to (\ref{sigmaea}) except that the lowest energy of the
Hamiltonian operator $H$ has been replaced by the lowest energy of the operator
$D_{\text{FMC}}$.
In conclusion the stability of the algorithm is related to the lowest eigenvalue
of the FMC diffusion operator, $\tilde{E}_0^B$.
It is clear from (\ref{fluctfmc}) that the higher this eigenvalue is, the more
stable the simulation will be.
In the next section, we will discuss the allowed values of $\tilde{E}_0^B$.
This will  prove that FMC is not a stable method in general, but
is more stable than any standard transient method.

\subsection{Stability of the Fermion Monte Carlo algorithm}
In this section we prove that the lowest eigenvalue of the FMC operator,
$\tilde{E}_0^B$, has the following upper and lower bounds 
\begin{eqnarray}
  \tilde{E}_0^B & \leq & E_0^F
 \label{stableineq}
  \\
  \tilde{E}_0^B & > & E_0^B
\label{improvstable}
\end{eqnarray}
>From the expression of the variance, Eq. (\ref{fluctfmc}), one can easily
understand the meaning of these two inequalities.
The first inequality indicates that FMC is not a stable
method, the stability being achieved only in the limit $\tilde{E}_0^B=E_0^F$.
Note that, even for very simple systems, this stability is in general not obtained.
This important point will be illustrated in the next section.
The second inequality shows that FMC is more stable than any standard 
transient DMC method (nodal release method). Indeed, the exponent associated with the explosion 
of fluctuations, Eq. (\ref{fluctfmc}), is smaller than in the standard case, Eq.(\ref{sigmaea}).
Before giving a mathematical proof of these two inequalities, let us first present
some intuitive arguments in their favor.
The first inequality, Eq. (\ref{stableineq}), takes its origin in the fact that the 
signal -the antisymmetric component of $f_t$, Eq.(\ref{aft}) is extracted from the population
of pairs of walkers, Eq.(\ref{f_tfmc}), and, consequently, cannot grow faster than the
population of pairs itself, Eq.(\ref{asympit}).
The second inequality can be understood as follows.
Without the cancellation process, the FMC method reduces to
two correlated DMC algorithms. The number of walkers grows as in a standard 
DMC, namely $ \sim e^{-(E_0^B-E_T)t}$. The cancellation
process obviously reduces the growth of the population of walkers, Eq.
(\ref{asympit}), and, thus, we should expect that $\tilde{E}_0^B  >  E_0^B$.
Now, let us give some more rigorous proofs. For that purpose we 
compare the Fermion Monte Carlo operators with and without cancellation process.
Without the cancellation process the FMC diffusion operator reads
  \begin{eqnarray}
 D -E_T & \equiv & {\psi_G^+}  (H^+-{E_L}^+) \frac{1}{{\psi_G}^+}
+{\psi_G}^-   (H^--{E_L}^-) \frac{1}{{\psi_G}^-} 
\label{langevin+-}
\\
& & - \frac{1}{2} \frac{\partial^2}{\partial {R^+}_\mu \partial {R^-}_\nu}[c_{\mu \nu} .]
\label{langevincorr}
 \\
& & + \text{max}({E_L}^+,{E_L}^-) -E_T
\label{branchingpair}
\\
& & + \frac{1}{2} [{E_L}^+-\text{max}({E_L}^+,{E_L}^-)]
e^{(P{\bf R}^+- {\bf R}^-).{\bf \nabla}^-. } 
\label{hoping-}
\\
& & + \frac{1}{2} [{E_L}^--\text{max}({E_L}^+,{E_L}^-)] 
e^{ (P{\bf R}^-- {\bf R}^+). {\bf \nabla}^+. }
\label{hoping+}
\end{eqnarray}
where the operators $H^\pm$ are both identical to $H$, except
that $H^+$ and  $H^-$ act on the space of positive and negative configurations, respectively.
\begin{equation}
H^{\pm} \equiv H ({\bf R}^{\pm}) 
= -\frac{1}{2} \sum_\mu \frac{\partial^2}{{\partial R^{\pm}_\mu}^2} + V ({\bf R}^\pm).
\end{equation}
The coefficients $c_{\mu\nu}$ are real coefficients and will be defined below. 
To justify that this operator is the diffusion operator
corresponding to FMC with no cancellation process, we need to 
check that the short-time dynamics described by
  \begin{equation}
\frac{\partial \Pi_t}{\partial t} = -( D-E_T) \Pi_t
\label{fmcwcan}
  \end{equation}
is indeed realized via the two first steps of the FMC algorithm (Langevin and
branching steps).
In the expression of $D$, the two operators appearing in Eq.(\ref{langevin+-})
define a Fokker Planck operator in analogy to Eq.(\ref{fkpp}).
This operator is the diffusion operator associated with the Langevin process,
Eq.(\ref{corrlanffmc}).
The term in Eq.(\ref{langevincorr}) is a coupling term between the moves of
positive and negative walkers taking into account the correlation of the
gaussian random variables $\eta_\mu^+$ and $\eta_\mu^-$.
The quantities $c_{\mu \nu} ({\bf R}^+,{\bf R}^-)$ introduced in Eq.(\ref{langevincorr}) 
are nothing but the covariance of these variables
\begin{equation}
c_{\mu \nu} ({\bf R}^+,{\bf R}^-) = \langle \eta_\mu^+ \eta_\nu^- \rangle.
\end{equation}
The three last contributions describe the branching processes  at work in FMC.
One recognizes in Eq.(\ref{branchingpair}) the branching of a pair, Eq.(\ref{wminpair}).
The two following contributions, Eqs.(\ref{hoping-}) and
(\ref{hoping+}), correspond to the
creation of pairs $({\bf R}^+,P{\bf R}^+)$ and $(P{\bf R}^-,{\bf
  R}^-)$, with the respective weights given by (\ref{wsingle1}) and (\ref{wsingle2}).
Note that in Eqs.(\ref{hoping-},\ref{hoping+}) the operator
$e^{ (P{\bf R}^--  {\bf R}^+).  {\bf \nabla}^+ }$ is written in a symbolic form 
representing a translation of the vector ${\bf R}^+$ to $P{\bf R}^-$, the action of this operator 
on the pair $({\bf R}^+,{\bf R}^-)$ being indeed to create the pair $(P{\bf R}^-,{\bf R}^-)$.

Now, let us prove that the lowest eigenvalue of $D$ is $E_0^B$ (bosonic ground-state).
For that purpose, it is convenient to introduce the operator ${\cal R}$ which transforms a
distribution of pairs of walkers into a distribution of walkers and then to define 
the following reduced density 
\begin{equation}
{\cal R} {\Pi}_t ({\bf R}) \equiv \int dR^\prime \left[\frac{{\Pi}_t ({\bf
      R},{\bf R}^\prime)}{\psi_G^+ ({\bf R})}
+\frac{{\Pi}_t ({\bf R}^\prime,P{\bf R})}{\psi_G^- ({\bf R})} \right].
\label{reduceddensity}
\end{equation}
The density ${\cal R} {\Pi}_t$ represents the sum of the distributions sampled by each type of 
walkers when the contribution of the other type of walkers is integrated out.
Using the explicit expression of $D$ it is a simple matter of algebra to
verify that
\begin{equation}
{\cal R}  D {\Pi_t}=  H {\cal R }{\Pi}_t.
\label{relevint}
\end{equation}
% MC: This relation may look very formal, but it has a physical  meaning.
% Si tu ne dis rien de plus, il faut enlever cette phrase.
Using Eqs.(\ref{relevint}) and (\ref{fmcwcan}), one can also write 
\begin{equation}
\frac{\partial {\cal R} \Pi_t}{\partial t} = - (H-E_T) {\cal R }\Pi_t
\label{evof}
\end{equation}
which means that the reduced  density evolves under the dynamics of $H$.
In other words, the set of positive and negative walkers sample the same 
distribution as a standard Diffusion Monte Carlo algorithm.
Now, suppose that $\lambda_S$ is the lowest eigenvalue of $D$ and $\Pi_S$ the corresponding 
eigenstate (the stationary density of the process described by Eq.(\ref{fmcwcan}))
\begin{equation}
D  \Pi_S = \lambda_S \Pi_S.
\end{equation}
Applying ${\cal R}$ on both sides of this identity and using the relation (\ref{relevint}) 
one gets
\begin{equation}
H {\cal R}  \Pi_S = \lambda_S {\cal R} \Pi_S.
\end{equation}
In other words, the reduced density ${\cal R} \Pi_S$ is a positive eigenstate of
$H$ with eigenvalue $\lambda_S$. The bosonic state being non-degenerate, we can conclude that
$\lambda_S=E_0^B$. This ends our proof.

Let us now consider the genuine FMC diffusion operator including the cancellation process,
$D_{FMC}$. To simplify the notations let us suppose that we are in the symmetric case for which 
$\psi_G^+=\psi_G^-=\psi_G$ (the common guiding function is symmetric under permutation 
of particles). This particular case is much simple because when walkers meet, there is a 
full cancellation and no residual branching. From an operatorial point of view the 
cancellation step consists in introducing a projection operator, $P_c$, 
at each step of the dynamics
\begin{eqnarray}
P_c \equiv  [1-\int dR \mid R^+=R, R^-=R \rangle \langle R^+=R, R^-=R\mid].
\label{p_c}
\end{eqnarray}
where $\mid R^+, R^- \rangle$ denotes the usual tensorial product.
The full FMC diffusion operator can thus be written as
\begin{equation}
D_{\text{FMC}} \equiv P_c  D.
\label{hprimedef}
\end{equation}
It is important to realize that the $D_{\text{FMC}}$ operator defined via $P_c$ and $D$
(Eqs.(\ref{p_c}) and (\ref{hoping+})) represents indeed an equivalent operatorial description of 
the stochastic rules of FMC described in section IV.B (Langevin, branching and cancellation
steps). Note also that using the expression
(\ref{hprimedef}) of $D_{\text{FMC}}$, we have a simple alternative way of 
recovering the proof just presented above that FMC is a bias-free approach. Since this is 
an important point of this work, let us present this alternative proof.
The action of the projection operator $P_c$ is to remove from the sample components 
of the form $\mid R^+=R, R^-=R \rangle$
for which the antisymmetric component of the reduced density is zero
\begin{equation}
{\cal A} {\cal R} \mid R^+=R, R^-=R \rangle 
= {\cal A} \frac{1}{\psi_G ({\bf R})} (\mid R \rangle + \mid P R \rangle)=0.
\label{p0part}
\end{equation}
In other words one has the following algebraic identity
\begin{equation}
{\cal A} {\cal R} P_c  = {\cal A} {\cal R}.
\label{p0gen}
\end{equation}
In the general case where $\psi_G^+ \ne \psi_G^-$, the cancellation procedure
still corresponds to define a new operator written as in Eq.(\ref{hprimedef}) with
$P_c$ satisfying the same identity as in (\ref{p0gen}).
Applying ${A\cal R}$ to the L.H.S. and R.H.S. of equation (\ref{evotilde}), one has
\begin{equation}
 \frac{  \partial {\cal A} {\cal R} \Pi_t} {\partial t} 
= - {{\cal A}\cal R}  D_{\text{FMC}}  \Pi_t = -{\cal A} (H-E_T) {\cal R} \Pi_t.
\end{equation}
This equation indicates that the evolutions of the antisymmetric component of the reduced
density under the dynamics of ${D}_{\text {FMC}}$ and $H$ are identical.
This confirms that the energy estimator, Eq.(\ref{f_tfmc}), or any observable estimator not 
coupling directly positive and negative walkers, is not biased. In the case of the 
energy, the estimator can be written as a function of the reduced density as follows
\begin{equation}
E_0^F = \frac{\langle \psi_T \mid H {\cal R}  {\Pi}_t \rangle}
{\langle \psi_T \mid {\cal R}  {\Pi}_t \rangle}.
\label{easymb}
\end{equation}

Let us now turn back to our discussion of the stability of FMC. For that, we need 
to compare the lowest eigenvalue $\tilde{E}_0^B$ of ${D}_{\text{FMC}}$ and, $E_0^B$, the lowest 
eigenvalue of $H$. Now, it is clear from the definition of ${D}_c$, Eq.(\ref{hprimedef}),
that the following relation holds
\begin{equation}
\tilde{E}_0^B > E_0^B.
\label{bettstab}
\end{equation}
Indeed, the action of $P_c$ present in the definition of 
${D}_c$, Eq.(\ref{hprimedef}), consists in removing positive coefficients within the
extradiagonal part of ${D}$. As well-known, a consequence of such a matrix (or operator) 
manipulation is to increase the energy of the lowest eigenvalue of the matrix.
Expressed in a more physical way, the cancellation process reduces the growth of the
population of pairs: $e^{-(\tilde{E}_0^B-E_T)t} < e^{-(E_0^B-E_T)t}$.

To summarize, we have shown that FMC reduces the instability of fermion simulations. 
The signal-over-noise ratio decreases as $e^{(\tilde{E}_0^B-E_0^F) t}$, 
where $\tilde{E}_0^B$ is the lowest eigenvalue of the FMC operator, ${D}_{\text{FMC}}$.
Because of the inequality (\ref{bettstab}), this ratio decreases slower in FMC that in any
standard transient DMC or nodal release methods.
We have shown that the cancellation process is at the origin of this improvement;
however, as we shall see in the next section, the cancellation process is
efficient (i.e., we have a small difference $\tilde{E}_0^B-E_0^F$) only if the correlation between 
walkers described by the coupling terms $c_{\mu\nu}$ is
introduced. This feature is important, particularly in high-dimensional spaces where the
probability of meeting and cancelling becomes extremely small for independent walkers.
As a result, the correlation of positive and negative walkers is a fundamental 
feature of FMC.
The quantitative effect of the correlation on the stabilization of the algorithm is not easy 
to study theoretically and to optimize in the general case.
In the next section we will give a numerical illustration, for a simple system,
of the interplay between cancellation and correlation (via the $c_{\mu\nu}$ parameters), 
and also of the role of the choice of the guiding functions, $\psi_G^+$ and $\psi_G^-$.

\section{Numerical study}

\subsection{The model: 2D-harmonic oscillator on a finite grid}
In this section we study the FMC method on a very simple model on a lattice.
For this model it is possible to calculate $E_0^F$ (fermionic ground-state energy), 
$E_0^B$ (bosonic ground-state energy), and $\tilde{E}_0^B$ the lowest eigenvalue of the FMC 
operator by a standard deterministic method (exact diagonalization).
The results obtained for this simple model will provide us with a well-grounded 
framework to interpret the Fermion Monte Carlo simulations.
The second motivation is that, using such a simple model, it is possible to study the limit 
of large number of walkers, large with respect to the dimension of the Hilbert space considered.
This possibility turns out to be essential to better understand the FMC algorithm.

Our model is based on the discretization of a system describing two-coupled harmonic oscillators
\begin{equation}
H = -\frac{1}{2} ( \frac{\partial^2}{\partial x^2}
+ \frac {\partial^2}{\partial y^2} ) + V(x,y)
\end{equation}
with
\begin{equation}
V(x,y) = \frac{1}{2} x^2 + \frac{1}{2}\lambda y^2 + xy
\end{equation}
In the following we shall take $\lambda=2$.
Now, we define the discretization of this model on a  $N$x$N$ regular grid
($N$ odd). A grid point ${\bf R}_i  \;\; i \in (1, \dots, N^2)$ has the following
coordinates 
\begin{equation}
{\bf R}_i  \equiv \left((-\frac{N}{2}+k-1)\delta_x, (-\frac{N}{2}+l-1)
\delta_y\right) \;\;\;  k \in [1 \dots N], \;\; l \in [1\dots N]
\end{equation}
where 
\begin{equation}
\delta_x=\delta_y=x_{max}/N
\label{xmax}
\end{equation}
On this lattice the Hamiltonian has a corresponding discrete representation given
by a finite matrix.
The diagonal part of the matrix reads
\begin{equation}
H_{ii}= \frac{1}{{\delta_x}^2} + \frac{1}{{\delta_y}^2} + V(R_i)   \;\; i \in (1,\dots,N^2)
\end{equation}
and the off-diagonal part reads
\begin{eqnarray}
H_{ij} & = &  -\frac{1}{2{\delta_x}^2}  \text{ \ \  when $R_i$ and $R_j$ are nearest-neighbors on 
the lattice} \nonumber \\ 
H_{ij} & = & 0 \text{ \ \   otherwise}
\end{eqnarray}

This Hamiltonian is symmetric with respect to the inversion $P$ of center $O = (0,0)$.
\begin{equation}
P (x,y) \equiv (-x,-y)
\end{equation}
As a consequence, the eigenfunctions are either symmetric or antisymmetric under
$P$. We are interested in the energy $E_0^F$ of the lowest antisymmetric eigenstate,
$\phi_0^F$. Even for this simple system, we are confronted with a sign
instability and a genuine ``sign problem''. Indeed, the sign of $\phi_0^F$ for each grid 
point cannot be entirely determined by symmetry. 
Symmetry implies only that $\phi_0^F$ vanishes at the inversion center and that the two-dimensional
pattern of positive and negative values for $\phi_0^F$ is symmetric by inversion. The precise 
delimitation between positive and negative zones of the wavefunction (analogous to 
``nodal surfaces'' for continuous systems) is not known.

Let us now introduce the trial functions $\psi_T$ and $\psi_S$. $\psi_T$ 
has to be an (antisymmetric) approximation of $\phi_0^F$ and $\psi_S$ some symmetric and positive
approximation of the lowest eigenstate, $\phi_0^B$.
We have chosen them as discretizations of the exact solutions of the initial continuous model. 
To find these solutions we perform a diagonalization of the quadratic form of the potential
\begin{equation}
V(x,y) = \frac{1}{2} x^2 + \frac{1}{2}\lambda y^2 + xy
 = \frac{1}{2} k_1 \tilde{x}^2 + \frac{1}{2}k_2 \tilde{y}^2.
\end{equation}
It is trivial to verify that
$$
k_1=\frac{\cos^2{\theta} -\lambda \sin^2{\theta}}{cos{2\theta}}
$$
$$
k_2=\frac{\lambda \cos^2{\theta} - \sin^2{\theta}}{cos{2\theta}}
$$
$$
\tan{2\theta}= \frac{2}{\lambda-1}
$$
with
\begin{equation}
\tilde{x}= x \cos{\theta} - y \sin{\theta}
\end{equation}
and
\begin{equation}
\tilde{y}= x \sin{\theta} + y \cos{\theta}.
\end{equation}
If $k_1 < k_2$ we choose as trial wavefunction
\begin{equation}
\psi_T= \tilde{x}
\exp{( -\frac{\sqrt{k_1}}{2} \tilde{x}^2 -\frac{\sqrt{k_2}}{2} \tilde{y}^2)},
\end{equation}
while, in the other case, we take 
\begin{equation}
\psi_T= \tilde{y}
\exp{( -\frac{\sqrt{k_1}}{2} \tilde{x}^2 -\frac{\sqrt{k_2}}{2} \tilde{y}^2)}
\end{equation}
The lowest (symmetric) eigenstate is chosen to be
\begin{equation}
\psi_S= 
\exp{(-\frac{\sqrt{k_1}}{2} \tilde{x}^2 -\frac{\sqrt{k_2}}{2} \tilde{y}^2)}.
\end{equation}
Note that, in the limit of a very large system the trial functions, $\psi_T$ and $\psi_S$ 
reduce to two exact eigenstates of $H$; however, this is not the case for finite systems.

%%%%%%%%%%%%%%%%%%%%%%%%%%%%%%%%%%%%%%%%%%%%%%%%%%%%%%%%%%%%%%%%%%%%%%%%%%%
\subsection{FMC on the lattice}
%%%%%%%%%%%%%%%%%%%%%%%%%%%%%%%%%%%%%%%%%%%%%%%%%%%%%%%%%%%%%%%%%%%%%%%%%%%
Before presenting our results, let us say a few words about 
the implementation of the FMC on the lattice.
The same ingredients as in the continuum case hold, except that in the lattice case
the Langevin process is realized through a discrete transition probability matrix.
The probability for a (positive or negative) walker $i$ to go to $j$, after a time step $\tau$ is 
\begin{equation}
P^{\pm}(i\rightarrow j) \equiv \frac{\psi_G^{\pm}({\bf R}_j)}{\psi_G^{\pm}({\bf R}_i)}
<{\bf R}_j|1-\tau(H-E_L^{\pm})|{\bf R}_i>
\label{Plattdef}
\end{equation}
where 
$\tau$ is small enough to have a positive density, namely
\begin{equation}
\tau < \frac{1}{\text{Max}[H_{ii}-E_L^\pm(i)]}.
\label{taumax}
\end{equation}
The local energies are defined as in the continuum, Eq.(\ref{elgw}) 
\begin{equation}
E_L^{\pm}({\bf R})= \frac{H \psi_G^{\pm}}{\psi_G^{\pm}}({\bf R})
\nonumber
\end{equation}
with the same expression for the guiding functions $\psi_G^{\pm}$, Eq.(\ref{psig+-def}) 
\begin{equation}
\psi_G^{\pm}({\bf R}) \equiv \sqrt{\psi_S^2 + c^2 \psi_T^2} \pm c\psi_T.
\label{psigpm}
\end{equation}
Let $(i_1,i_2)$ represents a given pair of positive and negative walkers
$({\bf R}^+_{i_1},{\bf R}^-_{i_2})$.
In a standard diffusion Monte Carlo (no correlation and no cancellation
of positive and negative walkers), the density of pairs of positive and negative
walkers $\Pi^{(k)}_{i_1i_2}$ evolves as follows in one time-step
\begin{equation}
\Pi^{(k+1)}_{i_1 i_2}= \sum_{j_1 j_2} \Pi^{(k)}_{j_1 j_2} P^+(j_1 \rightarrow
i_1) W^+_{j_1} P^-(j_2 \rightarrow i_2) W^-_{j_2}
\end{equation}
where $W^\pm$ is the Feynman-Kac weight, Eq.(\ref{FKW+-}).
To build the FMC algorithm, one first correlate the two stochastic processes 
$P^+$ and $P^-$, Eq.(\ref{Plattdef}).
The way it is performed here is the counterpart of the correlation term introduced 
by Liu {\sl et al.} \cite{Liu1994} in the continuum case, Eq.(\ref{cor+-}).
With such a choice,  
the positive and negative walkers of a pair tend to get closer or to move away 
in a concerted way.
For the lattice case it is done as follows.
The positive walker $j_1$ is connected by $P^+$ to a finite number of states
${j_1^c}$ (here, maximum five)
with probability $P^+ (j_1 \rightarrow j_1^c)$.
The negative walker $j_2$ is connected to a finite number of
states $j_2^c$ with the probability $P^- (j_2 \rightarrow j_2^c)$.
The states $j_1^c$ are ordered taking as criterium the distance to
the negative walker $j_2$, $|{\bf R}_{j_1^c}-{\bf R}_{j_2}|$.
We do the same for the states  $i_2^c$ ordered by their distance 
with respect to the positive walker.
An {\it unique} random number uniformly distributed between 0 and 1 is 
then drawn and the repartition functions of the two probability measures, 
$p (j_1^c) \equiv P^+ (j_1 \rightarrow j_1^c) $ 
and $p (j_2^c) \equiv P^- (j_2 \rightarrow j_2^c)$ are then sampled using this common 
random number.
The new pair $(j_1^c,j_2^c)$ is drawn accordingly.
Such a procedure defines a {\it correlated} transition probability in the space of pairs
\begin{equation}
P_c (j_1 j_2 \rightarrow i_1 i_2)  \ne P (j_1   \rightarrow i_2)
P  (j_1   \rightarrow i_2)
\label{corrmove}
\end{equation}
whose role is to enhance the probability of having positive and negative walkers 
meeting at the same site. Note that, by construction, the correlation introduced via 
$P_c$ does not 
change the individual (reduced) densities associated with each type of walker (positive/negative).
Now, let us write down explicitly the FMC rules in our lattice case, that is,
the one time-step ($k\rightarrow k+1$) evolution of the density of pairs $\Pi^{(k)}_{j_1 j_2}$:

(i) {\sl Correlation and branching.}
The branching of an individual pair $(j_1,j_2)$, Eq.(\ref{branchingpair}),
corresponds to the following evolution of the density
\begin{equation}
\Pi^{(k+1)}_{i_1 i_2}= \Pi^{(k)}_{i_1 i_2} + \sum_{j_1 j_2} \Pi^{(k)}_{j_1 j_2} 
P_c(j_1 j_2 \rightarrow i_1 i_2) \text{Min}[ W^+_{ i_1},
W^-_{i_2} ]
\label{eqfmc1}
\end{equation}
The creation of pairs, Eqs.(\ref{hoping-},\ref{hoping+}), can be written as follows
\begin{equation}
\Pi^{(k+1)}_{i_1 P(i_1)}= \Pi^{(k)}_{i_1 P(i_1)} + \sum_{j_1 j_2} \Pi^{(k)}_{j_1 j_2} P_c(j_1 j_2 
\rightarrow i_1 i_2) 
\theta[W^+_{i_1},W^-_{i_2}] |W^+_{i_1}-W^-_{i_2}|/2
\label{eqfmc2}
\end{equation}
\begin{equation}
\Pi^{(k+1)}_{P(i_2) i_2}= \Pi^{(k)}_{P(i_2) i_2} + 
\sum_{j_1 j_2} \Pi^{(k)}_{j_1 j_2} P_c(j_1 j_2 \rightarrow i_1 i_2)
(1- \theta[W^+_{i_1},W^-_{i_2}]) |W^+_{i_1}-W^-_{i_2}|/2
\label{eqfmc3}
\end{equation}
where $\theta(x,y)=1$  if $x > y$, $\theta(x,y)=0$, otherwise.

(ii) {\sl Cancellation.}
The cancellation process is done when a positive walker and a negative walker
meet $i_1=i_2=i$
\begin{equation}
\Pi^{(k+1)}_{ii} = \left[ 1 - \text{min} (\frac{\psi_G^+}{\psi_G^-} (i),
\frac{\psi_G^-}{\psi_G^+} (i)) \right] \Pi^{(k)}_{ii}.
\label{eqfmc4}
\end{equation}
If $\psi_G^+(i) > \psi_G^-(i)$ we have
\begin{equation}
\Pi^{(k+1)}_{P(i) i}= \Pi^{(k)}_{P(i) i} + \frac{[1-\psi_G^-(i)/\psi_G^+(i)]}{2} \Pi^{(k)}_{i i}.
\label{eqfmc5}
\end{equation}
If $\psi_G^+(i) < \psi_G^-(i)$ we have
\begin{equation}
\Pi^{(k+1)}_{i P(i)}= \Pi^{(k)}_{i P(i)} + \frac{[1-\psi_G^+(i)/\psi_G^-(i)]}{2} \Pi^{(k)}_{i i}.
\label{eqfmc6}
\end{equation}
The operations (\ref{eqfmc1},\ref{eqfmc2},\ref{eqfmc3},\ref{eqfmc4},\ref{eqfmc5},\ref{eqfmc6})
describe the one time-step dynamics of the simulation.
At iteration $k$, the distribution of pairs $\Pi^{(k)} (i_1,i_2)$ is obtained
and the transient estimator of the energy (\ref{eapair}) can be computed from 
\begin{equation}
E(k) = \frac{ \sum_{i_1,i_2} \Pi^{(k)}_{i_1,i_2} [ \frac{H\psi_T (i_1)}{\psi_G^+(i_1)}- \frac{H\psi_T (i_2)}{\psi_G^-(i_2)}]}
           { \sum_{i_1,i_2} \Pi^{(k)}_{i_1,i_2} 
[ \frac{\psi_T (i_1)}{\psi_G^+(i_1)}- \frac{\psi_T (i_2)}{\psi_G^-(i_2)}]}.
\label{energyest}
\end{equation}
This estimator converges to $E_0^F$ when $k \rightarrow \infty$.

Now, it is important to realize that the FMC rules just presented have, in principle, 
no stochastic character at all. For a finite system the FMC rules can be viewed 
as simple deterministic matrix manipulations between finite vectors of the Hilbert 
space (here, the multiplications to be performed have been explicitly written). 
At the beginning of the simulation (iteration $k=0$) 
some arbitrary starting vector $\Pi^{(0)}_{i_1,i_2}$ is chosen and, then, iterations 
are performed up to convergence. This important remark will allow us to organize our 
discussion of FMC results into two parts. In a first part (Section VI.C),  we perform explicitly 
the matrix multiplications involved and any stochastic aspect is removed from the results. 
Stated differently, this procedure can be viewed as performing a standard FMC simulation with an 
infinite number of walkers (the distribution at each point of the configuration space 
is exactly obtained, no statistical fluctuations are present). In a second part (Section VI.D), 
we implement the usual stochastic interpretation of the FMC rules using a finite number 
of walkers. This second part will allow to discuss the important consequences of the finiteness of 
the number of walkers and, in particular, the role played by the use of population 
control techniques.

\subsection{FMC using an infinite number of walkers: the deterministic approach}

\subsubsection{No systematic error: FMC is an exact method}

In this section we verify on our simple example that the FMC rules do not introduce any 
systematic error (bias). The energy expression (\ref{energyest}) has been computed by 
iterating the applications of the elementary 
operators defined by 
(\ref{eqfmc1},\ref{eqfmc2},\ref{eqfmc3},\ref{eqfmc4},\ref{eqfmc5},\ref{eqfmc6}). In practice, 
this corresponds to iterate a matrix ${G}_{\text{FMC}} (\tau)$.
The distribution $\Pi^{(k+1)}$ at iteration $k+1$ is obtained from $\Pi^{(k)}$ as follows
\begin{equation}
\Pi^{(k+1)} = {G}_{\text{FMC}} (\tau) \Pi^{(k)}
\end{equation}
The operator ${G}_{\text{FMC}} (\tau)$ has been applied a large number of times on some initial 
density $\Pi^{(0)}_{j_1 j_2}$ 
(a $N^4$ component vector, $N$ being the linear size of our lattice) and
the energy (\ref{energyest}) has been computed at each iteration $k$.
We have checked that the energy converges to the exact 
value, $E_0^F$, corresponding the lowest antisymmetric state, with all decimal places.
We have verified that this is true for several cases corresponding to $N$ ranging from 
4 to 17. For this specific problem these results confirm numerically that FMC is an exact method.

\subsubsection{Meeting time between positive and negative walkers}
Here, we want to illustrate quantitatively the fact that the correlation introduced 
via the probability transition helps greatly to lower the meeting time between 
positive and negative walkers. The influence of the choice of the guiding functions 
(here, parameter $c$ in Eq.(\ref{psigpm})) on the meeting time is also examined.
The meeting time is defined and evaluated as follows. 
We start with a configuration consisting of a 
positive walker located at a corner of the lattice and 
a negative walker located at the opposite corner. The positive and
negative walkers are moving stochastically with the transition probability
defined in (\ref{Plattdef}).
We test the two cases cases corresponding to uncorrelated and correlated moves.
The average time $\langle T \rangle$ 
(number of Monte Carlo steps times $\tau$) before the walkers meet is computed.
Our results are presented in Table \ref{table1} and are
given for different linear sizes of the grid. 
In this first case the guiding functions are chosen with a large antisymmetric component, $c=4$.
The results indicate clearly that more than one order of magnitude is gained by
correlating the moves of the two stochastic processes.
In Table \ref{table2} the same calculations are done, except that a symmetric
guiding function $c=0$ ($\psi_G^+=\psi_G^-=\psi_S$) is employed.
The average meeting time is found to be much lower than in the non-symmetric case, $c=4$, by 
nearly two orders of magnitude. This is true whether or not the stochastic processes are
correlated. This behaviour of the meeting time as a function of $c$ is not surprising. 
When $c$ is large the two functions
$\psi_G^+$ and $\psi_G^-$ are localized in the nodal pockets of $\psi_T$.
In the large-$c$ limit $\psi_G^+$ is zero whenever $\psi_T$ is negative
and $\psi_G^-$ is zero whenever $\psi_T$ is positive. In this limit the overlap between the two
distributions $\psi_G^+$ and $\psi_G^-$ is zero and we have a similar result for the 
probability that walkers meet.
>From these preliminary results the introduction of non-symmetric wavefunctions
seems to deteriorate the stability, this property will be confirmed in the next section.
\subsubsection{Stability in time of FMC}
We know from section V that the stability in time is directly related to the magnitude of 
the reduced Bose-Fermi energy gap
\begin{equation}
\tilde{\Delta}_{B-F} \equiv E_0^F -\tilde{E}_0^B
\label{gapdelta}
\end{equation}
where $\tilde{E}_0^B$ is the lowest eigenvalue of the FMC diffusion operator.
The greater this gap is, the faster the signal-over-noise ratio of the Monte Carlo simulation 
deteriorates, the full stability being obtained only when this gap vanishes.
The ultimate goal of an efficient FMC algorithm is to reduce the Bose-Fermi gap from its bare 
value $\Delta_{B-F}= E_0^F-E_0^B$ to a value very close to zero (ideally, zero).
The energies $\tilde{E}_0^B$ and $E_0^F$ can be calculated by exact diagonalization 
of the Fermion Monte Carlo operator, $G_{\text{FMC}}$.
In practice, we have chosen here to extract this information from large-time behaviour of the 
denominator of the energy, Eq. (\ref{energyest}).
In this regime 
the denominator behaves as in Eq. (\ref{denomfmcbehaviour}) where the reference
energy is adjusted to keep the number of pairs constant, $E_T=\tilde{E}_0^B$.
\begin{equation}
\langle {\cal D}(t=k\tau) \rangle \propto e^{-(E_0^F-\tilde{E}_0^B)t}
\label{Dbehavesim}
\end{equation}
The gap $E_0^F-\tilde{E}_0^B$ has been extracted from the large-k values of ${\cal D}(t=k\tau)$,
a quantity calculated deterministically by iterating the matrix $G_{\text{FMC}}$.
The results for different values of $c$ are reported in Table \ref{gapf(c)}.
For both the correlated and uncorrelated processes it is found that the gap
increases with $c$.
The minimal gap is obtained for $c=0$, that is when the guiding functions are symmetric
$\psi_G^+=\psi_G^-=\psi_S$.
This result is easily explained from the fact that there are two factors which favour 
the cancellation of walkers.
First, as we have already seen, the average meeting time is minimal when $c=0$ since, 
in this case, the overlap between the functions $\psi_G^+$ and $\psi_G^-$ is maximal. Furthermore, 
a full cancellation between the walkers is precisely obtained when $c=0$.
In conclusion, the greatest stability is obtained for a symmetric guiding function.

In Table \ref{gapf(N)} the gaps obtained at $c=0$ for different linear sizes $N$ are reported.
This table shows that, in the limit of large grids, that is for a system
close to the continuous model, the FMC algorithm reduces the gap by a factor $\sim 20$.
Such a result corresponds to a huge gain in the
stability since projection times about twenty times larger than in a standard nodal 
release method can be used.
  
\subsection{FMC using a finite number of walkers: the stochastic approach}

In the previous section the Fermion Monte Carlo method has been discussed and implemented 
by manipulating the exact Fermion Monte Carlo diffusion operator without making reference to 
any stochastic aspect (as already mentioned it is formally equivalent to use an infinite 
number of walkers).
Of course, for non-trivial systems it is not possible to propagate exactly 
the dynamics of the FMC operator. Accordingly, a finite population of walkers is introduced and 
specific stochastic rules allowing to simulate in average the action of the FMC operator are 
defined. Now, the important point is that in practice -like in any DMC method- one does not 
sample exactly the dynamics of the FMC operator because 
of the population control step needed to keep the finite number of walkers roughly constant.
\cite{umrigarJCP93,sorella98rec,khelif2000}
This step introduces a small modification of the sampled diffusion operator
which is at the origin of 
a systematic error known as the population control error. For a bosonic system, the error 
on the ground-state energy 
behaves as $\frac{1}{M}$ ($M$ is the average size of the population) and an 
extrapolation in $\frac{1}{M}$ can be done to obtain the exact energy.
For a fermionic system, as we shall see below, this behaviour is qualitatively different and, 
furthermore, depends 
on the guiding function used. To have a precise estimate of the mathematical behaviour of 
the population control error is fundamental since, in practice,
it is essential to be able to reach the exact fermi result using a reasonable number of walkers.
As we shall see later, this will not be in general possible with FMC.

In this section the Fermion Monte Carlo simulations are performed 
using Eqs.(\ref{eqfmc1},\ref{eqfmc2},\ref{eqfmc3},\ref{eqfmc4},\ref{eqfmc5}) which  
allow to propagate stochastically a population of $M$ walkers.
The population is kept constant during the simulation by using the 
stochastic reconfiguration Monte Carlo (SRMC) method.\cite{sorella98rec,khelif2000} 
In short, the SRMC method is a DMC method in which a reconfiguration step replaces the 
branching step. A configuration step
consists in drawing $M$ new walkers among the $M$ previous ones 
according to their respective Feynman-Kac weight (for the details, see the references 
given above).

\begin{center}
\begin{figure}[htp]
\includegraphics[height=12cm,width=12cm,angle=-90]{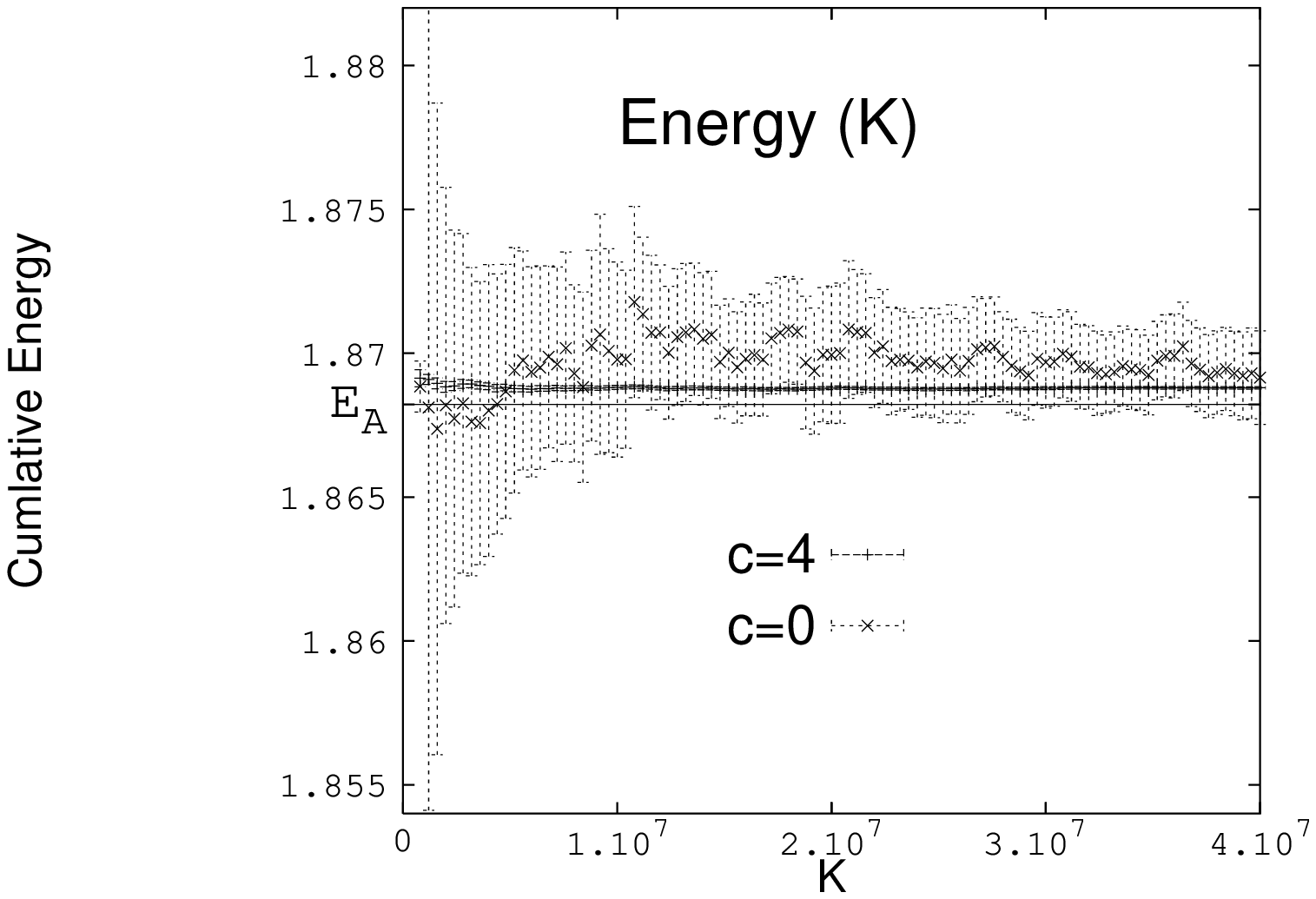}
\caption{$N=3$ Energy estimator as a function of the projection time
(number of iterations $K$). Comparaison between $c=0$ (large fluctuations) and
$c=4$ (small fluctuations).
The exact energy is $E_0^F=1.86822..$. Number of walkers
$M=100$. Number of Monte Carlo steps: $4.10^7$.
}
\label{fig1}
\end{figure}
\end{center}

\begin{center}
\begin{figure}[htp]
\includegraphics[height=12cm,width=12cm,angle=-90]{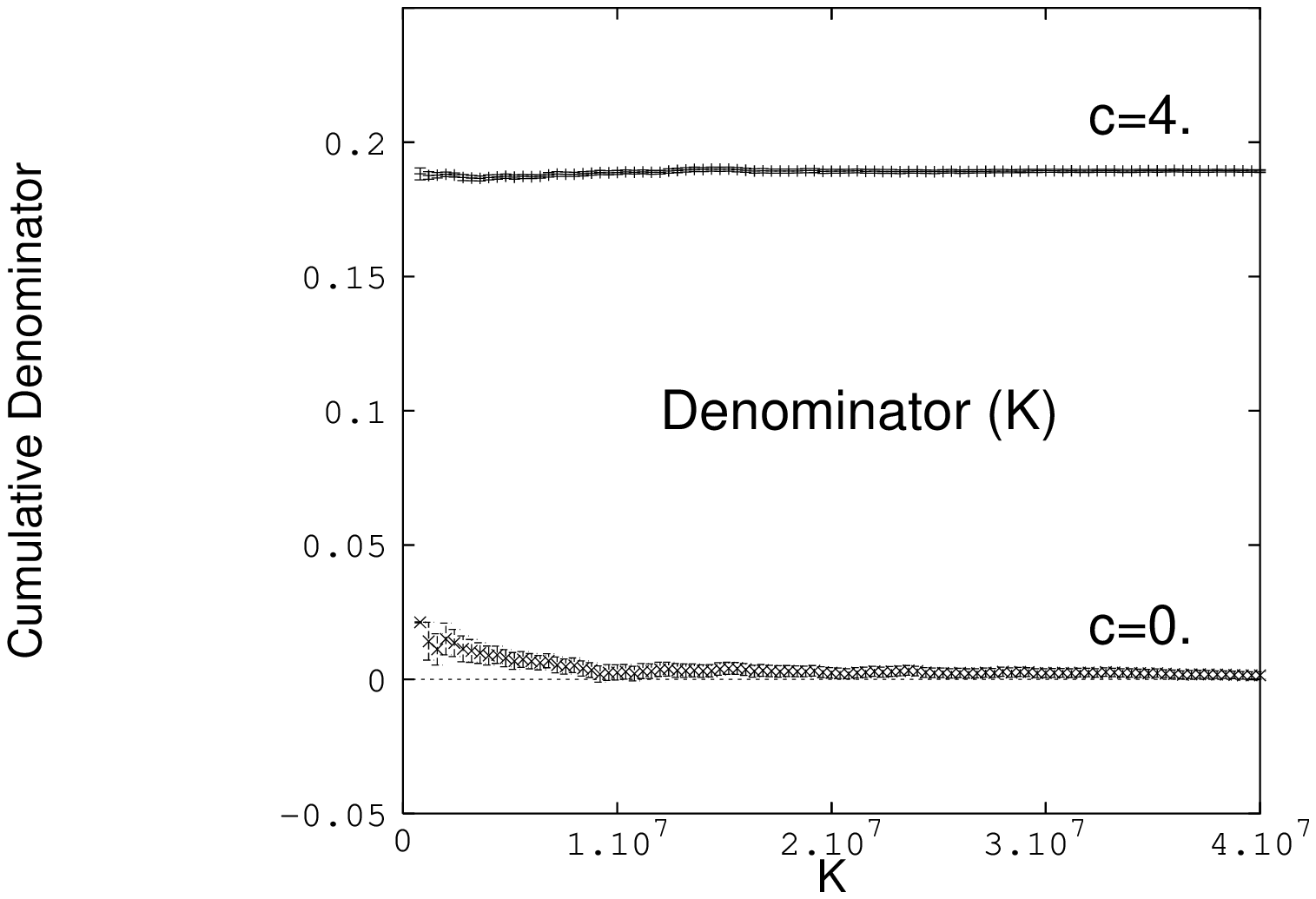}
\caption{$N=3$ Denominator as a function of the projection time (number of
iterations $K$). Comparaison between $c=4$ (upper curve) and
$c=0$ (lower curve).  Number of walkers
$M=100$. Number of Monte Carlo steps: $4.10^7$. }
\label{fig2}
\end{figure}
\end{center}

In Figure \ref{fig1} the time-averaged energy defined as
\begin{equation}
E(K) \equiv \frac{\sum_{k=1}^K {\cal N} (k)}{\sum_{k=1}^K {\cal D}(k)}
\end{equation}
is ploted as a function of $K$. In this formula
${\cal N}(k)$ and ${\cal D}(k)$ represent the numerator and denominator at 
iteration $k$ of the estimator (\ref{efmcN/D}) evaluated as an average over the population of 
pairs.
In Figure \ref{fig2} we plot the time-averaged denominator given by
\begin{equation}
{\cal D}_K =  \frac{1}{K} \sum_{k=1}^K {\cal D}(k).
\label{cumdenom}
\end{equation}
The time dependence of this quantity is interesting since it can be used as a measure 
of the stability of the algorithm.\cite{kalosfermion00} As we have shown above, 
the algorithm is stable only when the reduced Bose-Fermi energy gap, $\tilde{\Delta}_{B-F}
=E_0^F-\tilde{E}_0^B$ 
is equal to zero. Equivalently, the denominator (\ref{cumdenom}) must converge to a constant 
different from zero. In our simulations the number of walkers was chosen to be $M=100$, 
a value which is much larger than the total number of states of the system (here, nine states).
Of course, such a study is possible only for very simple systems. 
As seen on the Figures \ref{fig1},\ref{fig2}, and \ref{fig3} the results
obtained in the case $c=0$ (symmetric guiding function) and $c\neq 0$ are qualitatively 
very different.
Figure \ref{fig1} shows that, within statistical error bars, there is no systematic 
error on the energy when a symmetric guiding function is used, $c=0$.
However, the price to pay is that the statistical fluctuations are very large. 
This point can be easily understood by looking at the behaviour of the denominator,
Fig.\ref{fig2}. Indeed, this denominator vanishes at large times, thus
indicating that the simulation is not stable.
In sharp contrast, for $c=4$ (non-symmetric guiding functions), the 
statistical fluctuations are much more smaller (by a factor of about 40) 
but a systematic error appears for the energy. 
Furthermore, the denominator ploted in Fig.\ref{fig2} is seen to converge 
to a finite value. The stability observed in the case $c=4$ seems to confirm 
the results of Kalos {\sl et al.}\cite{kalosfermion00} for non-symmetric guiding 
functions ($c\ne 0$). However, the situation deserves a closer look. 
Indeed, the existence of this finite asymptotic 
value seems to be in contradiction with our theoretical analysis: the denominator should converge
exponentially fast to zero, and the algorithm should not be stable ($\tilde{\Delta}_{B-F} > 0$).
In fact, as we shall show now, the asymptotic value obtained for $c=4$ and the 
corresponding  stability result from a population control error.  
To illustrate this point, we have computed the average denominator as a function of the
population size $M$. Results are reported in Fig.\ref{fig3}. On this plot
we compare the population dependence of the denominator (\ref{cumdenom}) for $c=4$ and 
for a much smaller value of $c=0.5$. The values of $M$ range from $M=100$ to $M=6400$.
A first remark is that the population control error can be quite large
and is much larger for $c=4$ than for $c=0.5$.
In the Appendix it is shown that the theoretical asymptotic behaviour of the 
error as a function of $M$ is expected to be in $1/M$.
In the $c=0.5$ case, the denominator is clearly seen to extrapolate to zero like $\frac{1}{M}$
In the $c=4$ case, we can just say that the data are compatible
with such a behaviour but even for the largest $M$ reported in Fig.\ref{fig3} ($M=6400$)
this asymptotic regime is not yet reached. Much larger populations would be necessary.
This result illustrates the great difficulty in reaching
the asymptotic regime, even for such a simple system having only nine states.
Stated differently, the stability observed when using non-symmetric guiding
functions disappears for a large number of walkers, thus confirming that the
stability obtained at finite $M$ is a control population artefact.
Note that a large population control error on the denominator is not surprising.
Indeed, when $c\ne 0$, the local energies of the guiding functions, Eq.(\ref{elgw}),
have strong fluctuations because $\psi_G^\pm$ is far from any eigenstate of $H$ ($\psi_G^\pm$ 
contains a symmetric and an antisymmetric components).
In the case of a symmetric guiding function ($c=0$), the distribution of walkers is 
also symmetric at large times and, thus, the average of this distribution on the antisymmetric 
wavefunction $\psi_T$ must necessarily be zero. Consequently, in the $c=0$ case there is no 
control population error on the denominator, Fig.\ref{fig2}.

An example of the behaviour of the energy as a function of the population size $M$
is presented in figure \ref{fig4}. 
Some theoretical estimates of the energy bias dependence on $M$ are derived in the Appendix.
Let us summarize the results obtained.
When $c=0$ (use of a symmetric guiding function), 
the systematic error behaves as in a standard DMC calculation for a bosonic system
\begin{equation}
\delta E \propto \frac{1}{M}.
\end{equation}
However, the statistical fluctuations are exponentially large since the calculation is 
no longer stable.
Now, when $c > 0$ the systematic error has a radically different behaviour.
For a population size $M$ the control population error grows exponentially as a function 
of the projection time $t$
\begin{equation}
\delta E \propto \frac{1}{M}e^{t \tilde{\Delta}_{B-F}},
\label{formx}
\end{equation}
where $\tilde{\Delta}_{B-F}$ is the reduced Bose-Fermi energy gap. 
This dependence of the control population error as a function of the projection 
time is of course pathological and is a direct consequence of the use of non-symmetric 
guiding functions. Now, because of the form (\ref{formx}) it is clear that a population size 
exponentially larger than the projection time is necessary to remove the systematic population 
control error.
In practical calculations, for a given population of walkers $M$, one has to choose a finite 
projection time, $t$. This time has to be small enough to have a small finite population control 
error but, at the same time, large enough to extract the exact fermionic groundstate
from the initial distribution of walkers.
The best compromise is easily calculated and leads to the following expression of the
systematic error as a function of the number of walkers (for more details, see the Appendix)
\begin{equation}
\delta E \propto \frac{1}{M^{\gamma}}
\label{ebehav}
\end{equation}
where 
\begin{equation}
\gamma \equiv \frac{ \Delta_F }{ \tilde{\Delta}_{B-F} + \Delta_F} < 1
\label{gammabehav}
\end{equation}
and $\Delta_F$ is the usual Fermi gap 
(energy difference between the two lowest fermionic states).

In figure \ref{fig4} some numerical results for the $c=4$ 
and $c=0.5$ cases are presented. 
The number of walkers considered are $M=100,200,400,800$, and $M=1600$. 
No data are shown for the symmetric case, $c=0$, because of the very large 
error bars, see Fig.\ref{fig1}. The calculations have been done for the smallest 
system, $N=3$ (recall that the finite configuration space consists of 
only nine states) and for very large numbers of Monte Carlo steps (more than $10^8$).
As it should be, the systematic errors
are found to be larger for the $c=4$ case than for the $c=0.5$ case (note that the 
data corresponding to 
$M=800$ and $M=1600$ must not be considered because of their large statistical noise). 
The concavity of both curves confirms our theoretical result, 
$\gamma < 1$, Eqs.(\ref{ebehav},\ref{gammabehav}). However, it is clear that to get a quantitative 
estimate of this exponent is hopeless because of the rapid increase of error bars
as a function of $M$. The only qualitative conclusion which can be drawn by looking at the 
curves is that $\gamma_{c=4} < \gamma_{c=0.5}$, 
in agreement with our formula, Eq.(\ref{gammabehav}). Finally, let us insist on the fact
that, despite these very intensive calculations for a nine-state configuration 
space, no controlled extrapolation to the exact energy is possible. 

To summarize, when $\psi_G$ has an antisymmetric component, the error is expected 
to decrease -for $M$ large enough- very slowly as a function of the 
population size [algebraically with a (very) small exponent], 
while in the symmetric case the bias has a much more interesting $\frac{1}{M}$-behaviour.
However, in this latter case the price to pay is the presence of an exponential growth
of the statistical error. In both cases, and this is the fundamental point,
the number of walkers needed to get a given accuracy grows pathologically.
In addition, as illustrated by our data for the very simple model problem treated
here, the asymtotic regimes corresponding to the $1/M^\gamma$-behaviour appear
to be very difficult to reach in practice (very large values of $M$ are needed).

\section{Conclusion and perspectives}
The FMC method differs from the DMC method by correlating the diffusion of the walkers
and introducing a cancellation procedure between positive and negative walkers whenever they 
meet. In this work we have shown that the Fermion Monte Carlo approach is exact but 
in general not stable. FMC can be viewed as belonging to the class of transient DMC methods, 
the most famous one being probably the nodal release approach.
\cite{ceperleyalderPRL80,ceperleyalderPRL84}
However, in contrast with the standard transient methods, FMC 
allows to reduce in a systematic way the fermi instability.
The importance of this instability is directly related to the magnitude of some
``effective'' Bose-Fermi energy gap, $\tilde{\Delta}_{B-F}=E_0^F-\tilde{E}_0^B$, where 
$E_0^F$ is the exact Fermi energy and $\tilde{E}_0^B$ some effective Bose energy.
We have seen that this gap is intimately connected to the
the cancellation rate, that is to say, to the speed at which positive and negative walkers cancel.
We have shown that $E_0^B < \tilde{E}_0^B < E_0^F$, where $E_0^B$ is 
the standard bosonic ground-state 
energy. As an important consequence, the closest $\tilde{E}_0^B$ is from the exact fermionic energy,
the smoother the sign problem is. For the toy model considered, the reduction obtained for 
the instability is very large (orders of magnitude).
For large dimensional systems, there are strong indications in favor also of an 
important reduction. A first argument is purely theoretical. In FMC the walkers within a pair 
$({\bf R}_i^+,{\bf R}_i^-)$ are correlated in a such a way that ${\bf R}_i^+-{\bf R}_i^-$ 
makes a random move only in one dimension.
As a result there is a high probability that the walkers meet in a finite time 
even if they move in a high-dimensional space.
The second argument is numerical. As shown by previous authors, the impact of 
correlating walkers on the average meeting time is important even for 
much larger systems.\cite{kalosfermion00,CollettiFMC2005}

We have shown that the recent introduction of nonsymmetric guiding functions in 
FMC introduces a large systematic error which goes to zero very slowly
as a function of the population size [$\sim 1/M^\gamma$, $\gamma =
\Delta_F/(\tilde{\Delta}_{B-F} + \Delta_F)$ and $\Delta_F= E_1^F -E_0^F$ is the usual 
fermionic gap].
For an infinite number of walkers, this systematic 
error is removed and the algorithm recovers the Fermi instability.
Morever, we have shown that using such guiding functions does not in general improve the
stability. For a large enough number of walkers, the
simulation can be less stable than the simulation using a symmetric guiding function.
Finally, it is important to emphasize that the conclusion of this work is that the 
FMC algorithm is not a solution 
to the sign problem. However, it is a promising way toward ``improved'' transient methods. 
As a transient method, FMC is expected to converge much better than a standard nodal 
release method. We are presently working in this direction.

\section*{ACKNOWLEDGMENTS}
The authors warmly thank Malvin Kalos and Francesco Pederiva for
numerous useful discussions and for having inspired us the present work.
This work was supported by the ``Centre National de la Recherche
Scientifique'' (CNRS), Universit\'e Pierre et Marie Curie (Paris VI), Universit\'e 
Pauls Sabatier (Toulouse III), and Universit\'e Denis Diderot (Paris VII).
Finally, we would like to acknowledge computational support from IDRIS (CNRS, Orsay) and
CALMIP (Toulouse).

\appendix
\section*{APPENDIX: Population control error in FMC}
FMC, like any Monte Carlo method using a branching process, 
suffers from a so-called population control bias. This systematic error appears
because the branching rules (creation/annihilation of walkers) are implemented using a 
population consisting of a large but {\it finite} number of walkers. Nothing preventing
the population size from implosing or exploding, a population control step is required 
to keep the average number of walkers finite.
A standard strategy to cope with this difficulty consists in introducing a 
time-dependent reference energy whose effect is to slightly modify the elementary weights 
of each walker by a common multiplicative factor (close to one)
so that the total weight of the population remains nearly constant during the 
simulation. This step, which introduces some correlation between walkers and, therefore,
slightly modifies the stationary density, must be performed 
very smoothly to keep the population control error as small as possible.
In practice, for standard DMC calculations done with accurate trial wavefunctions
and population sizes large enough, the error
is found to be very small, in general much smaller than the statistical fluctuations. 
As a consequence, the presence of a population control bias is usually not considered as critical.
Here, the situation is rather different. 
In FMC the use of bosonic-type guiding functions introduces very large 
fluctuations of the local energy and the cancellation rules a very small signal-over-noise 
ratio for fermionic properties. In this case, it is not clear whether
the bias can be kept small with a reasonable number of walkers.

In this section we present an estimate of the population control bias in FMC. As we shall 
see our estimate shows that the sign problem is actually not solved but attenuated in FMC 
(an {\it exponentially} large number of walkers is needed to maintain
a constant bias as the number of electrons is increased).
The derivation presented in this section is very
general: it is valid for any exact fermion QMC method based on the use of a nodeless bosonic-type 
reference process and some projection to extract the Fermi ground-state. Accordingly,
we have chosen not to use the specific framework and notations of FMC but, 
instead, notations of a general DMC algorithm (transient method). Of course, we do not 
need FMC to introduce nonsymmetric guiding functions.
  The adaptation of what follows to FMC is straightforward.

\begin{center}
\begin{figure}[htp]
\includegraphics[height=12cm,width=12cm,angle=-90]{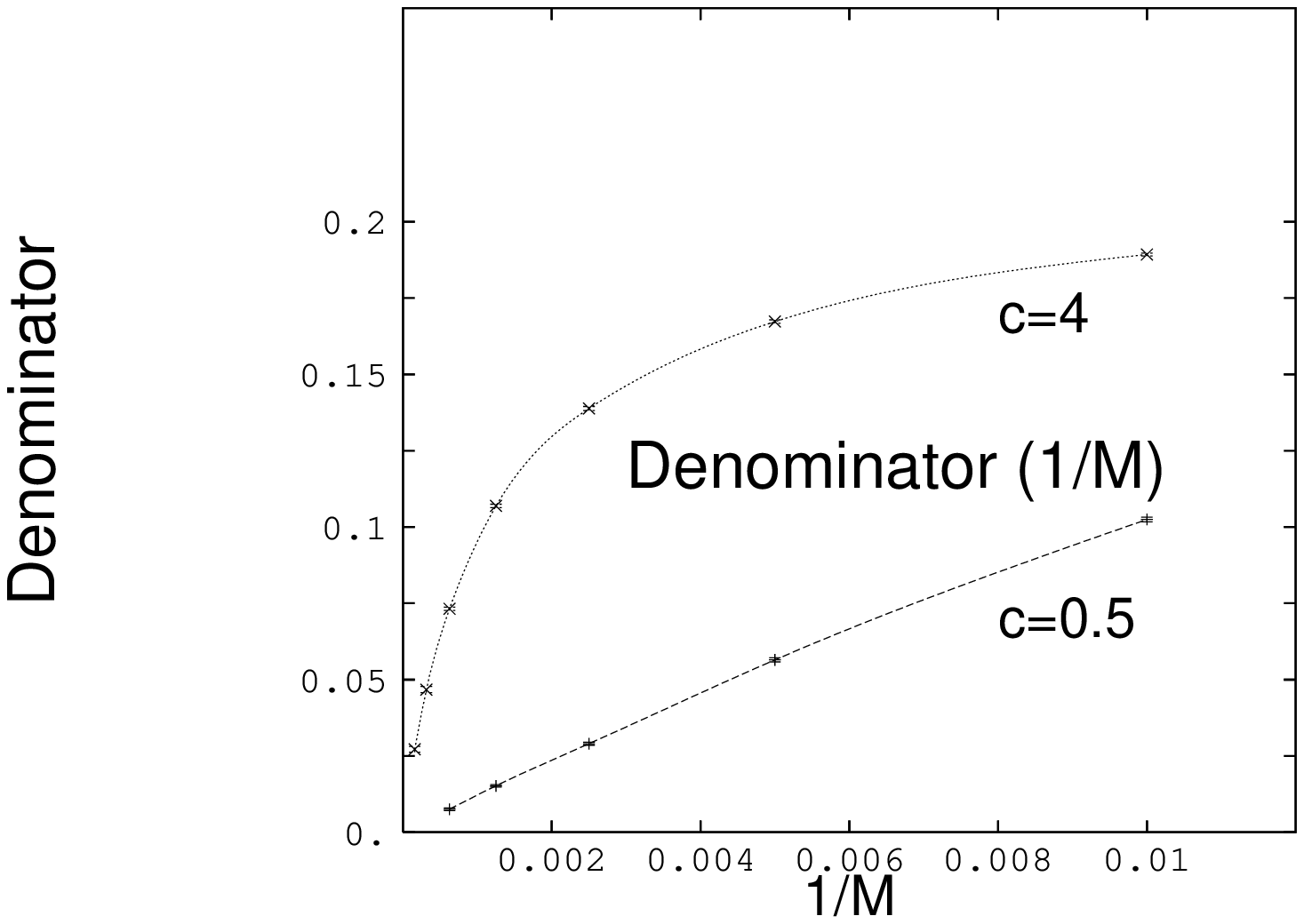}
\caption{$N=3$, time-averaged denominator, Eq.(\ref{cumdenom}), as a function of $1/M$. }
\label{fig3}
\end{figure}
\end{center}

In quantum Monte Carlo we evaluate stochastically 
the following expression for the lowest eigenstate of energy $E_0^F$
\begin{equation}
E_F^t = \frac{ \langle \psi_T \mid H e^{-t(H-E_T)} f_0 \rangle}
{\langle \psi_T \mid  e^{-t(H-E_T)} f_0 \rangle}
\label{energt}
\end{equation}
where $f_0$ is some positive initial distribution and $\psi_T$ an approximation of 
the eigenstate with energy $E_0^F$.
Here, we deal with a fermionic problem so $\psi_T$ must be antisymmetric.
The expression (\ref{energt}) gives the exact energy $E_0^F$  only when taking the limit $t \to \infty$. In practice for a finite $t$, there is a systematic error \begin{equation}
\Delta E_F^t \equiv E_F^t-E_0^F \propto e^{-\Delta_F t} \equiv e^{-(E_1^F-E_0^F) t}
\label{deltaeat}
\end{equation}
where $E_1^F$ is the energy of the first excited state in the antisymmetric sector and,
$\Delta_F$, the fermionic energy gap.
For an exact algorithm with one walker (e.g. Pure Diffusion Monte Carlo, 
\cite{Cafclav188,Cafclav288}) one computes the R.H.S of (\ref{energt}) by evaluating the following expression
\begin{equation}
E_F^t = \frac{\left\langle \frac{H \psi_T}{\psi_G} [R(t)] e^{-\int_0^t ds (E_L[R(s)]-E_T)} 
\right\rangle}
{\left\langle \frac{ \psi_T}{\psi_G} [R(t)] e^{-\int_0^t ds (E_L[R(s)]-E_T) }
\right\rangle}
\label{eexact}
\end{equation}
where $\psi_G$  is the guiding function, strictly positive, with eventually an antisymmetric component. 
We have also introduced in this expression, the local energy of the guiding function
\begin{equation}
E_L = \frac{H\psi_G}{\psi_G}
\end{equation} 
The integral  in (\ref{eexact}) is done  over the drifted random walks going from $0$ to $t$.
To simplify the notations we note (\ref{eexact}) as follows
\begin{equation}
E_F^t = \frac{\left\langle h^t  W^t \right\rangle}
{\left\langle p^t W^t \right\rangle}
\end{equation}
where 
\begin{eqnarray}
h^t & \equiv & \frac{H \psi_T}{\psi_G} [R(t)] \\
W^t & \equiv &  e^{-\int_0^t ds E_L[R(s)]} \\
p^t & \equiv & \frac{\psi_T}{\psi_G} [R(t)]
\end{eqnarray}
For $M$  independant walkers $R_i$  one has
\begin{equation}
E_F^t = \frac{\left\langle \frac{1}{M} \sum_i h_i^t W_i^t \right\rangle}
{\left\langle \frac{1}{M} \sum_i p_i W_i^t \right\rangle}
\label{eareconf}
\end{equation}
In the analysis presented here based on a population of $M$ the walkers branched 
according to their relative multiplicities, (the $M$ walkers are therefore no longer independant), one replace the individual weight $W_i$ by a global weight
\cite{khelif2000}
\begin{equation}
\bar{W}^t = \frac{1}{M}\sum_i W_i^t
\end{equation}
As a result the energy may be written as
\begin{equation}
E_F^t = \frac{\left\langle  \bar{h}^t \bar{W}^t \right\rangle}
{\left\langle \bar{p}^t \bar{W}^t \right\rangle}
\label{reconf}
\end{equation}
where
\begin{eqnarray}
\bar{h}^t & \equiv & \frac{1}{M}\sum_i h_i^t \\
\bar{p}^t & \equiv & \frac{1}{M}\sum_i p_i^t 
\end{eqnarray}
Expression (\ref{reconf}) is exact when the weights $\bar{W}^t$ are included. 
A control population error arises when one does not take into account the weights 
in expression (\ref{reconf}). This population control error is thus given by 
\begin{eqnarray}
\Delta E_F^M & = &  \frac{\left\langle  \bar{h}^t \right\rangle}
{\left\langle \bar{p}^t  \right\rangle}
-   \frac{\left\langle  \bar{h}^t \bar{W}^t \right\rangle}
{\left\langle \bar{p}^t \bar{W}^t \right\rangle}
 \\
   & = &  \frac{\left\langle  \bar{h}^t \right\rangle}
{\left\langle \bar{p}^t  \right\rangle} -
 \frac{\text{cov} ( \bar{h}^t, \bar{W}^t) + \left\langle \bar{h}^t\right\rangle\left\langle \bar{W}^t\right\rangle}
{\text{cov}( \bar{p}^t, \bar{W}^t)  +  \left\langle \bar{p}^t\right\rangle \left\langle \bar{W}^t\right\rangle }
\end{eqnarray}
or, after normalizing $\bar{W}^t$ in such a way that $ \langle \bar{W}^t\rangle =1$ 
(for example, by suitably adjusting the reference energy $E_T$),
\begin{equation}
\Delta E_F^M = \frac{\left\langle  \bar{h}^t \right\rangle}
{\left\langle \bar{p}^t  \right\rangle}
-  \frac{\text{cov} ( \bar{h}^t, \bar{W}^t) + <\bar{h}^t>}
{\text{cov}( \bar{p}^t, \bar{W}^t)  +  <\bar{p}^t> }
\label{eamcov}
\end{equation}
This is our basic formula expressing the systematic error at time $t$
resulting from the use of a finite population. Now, let us evaluate this expression 
in the large time $t$ and large $M$ regimes.
First, we consider the two denominators appearing in the R.H.S. of Eq.(\ref{eamcov}). 
Let us begin with the denominator of the second ratio:
\begin{equation} 
D_e \equiv \text{cov}( \bar{p}^t, \bar{W}^t) + <\bar{p}^t>
\label{D2def}
\end{equation}
Because this denominator is nothing but the denominator of the R.H.S. of Eq.(\ref{eexact})
we can conclude that $D_e$ does not depend on the population size $M$ and that
it vanishes exponentially fast
\begin{equation}
D_e =  K_e  e^{-(E_0^F-E_T)t},
\label{D2t}
\end{equation}
where $E_T$ is the reference energy, $E_T = E_0^B$ for a nodal release-type  method, 
and $E_T = \tilde{E}_0^B > E_0^B$ for the FMC method.
Let us now look at the other denominator of Eq.(\ref{eamcov})
\begin{equation}
D_a \equiv \langle \bar{p}^t\rangle.
\label{D1def}
\end{equation}
This denominator is the usual quantity evaluated during the simulation. It is 
an approximate quantity since it does not include the corrective weights.
The asymptotic behaviour of $D_a$ depends on $\psi_G$. We distinguish two cases:

(i) If $\psi_G$ is symmetric ($c=0$), the stationary density ($t$ large enough) is symmetric. 
Consequently, $D_a$, which is the average of an antisymmetric function, converges to zero 
exponentially fast at large times. For $M$ large enough, in a regime where the dynamics  
is close to the exact dynamics of the Hamiltonian, we know that the convergence is 
given by
\begin{equation}
D_a = K_a(M) e^{-\tilde{\Delta}_{B-F} t},
\label{D1t}
\end{equation}
where the coefficient $K_a$ depends on $M$ in general. This coefficient will be determined later.
>From equations (\ref{D1t}) and (\ref{D2t}), one can evaluate the error on the denominator
\begin{equation}
D_a-D_e = (K_a (M)-K_e) e^{-\tilde{\Delta}_{B-F} t}.
\label{difD1}
\end{equation}
We also know from the definitions of $D_a$ and $D_e$[(\ref{D1def}), (\ref{D2def})] that
the difference $D_a -D_e$ is a covariance of two averages
\begin{equation}
D_a -D_e = \text{cov}( \bar{p}^t, \bar{W}^t) 
\end{equation}
which, due to the central-limit theorem, behaves as
\begin{equation}
D_a -D_e  = \frac{1}{M} {\cal C} (t)
\label{difD2}
\end{equation}
where ${\cal C} (t)$ is some function of $t$.
Identifying (\ref{difD1}) and (\ref{difD2}), one finally obtains a determination of $K_a$.
Finally, we obtain the following behaviour for the systematic error on the denominator 
\begin{equation}
D_a -D_e = \text{cov}( \bar{p}^t, \bar{W}^t) \propto \frac{1}{M} e^{-\tilde{\Delta}_{B-F} t}
\end{equation}
(ii) If $\psi_G$ is not symmetric ($c\ne 0$) the stationary density has an antisymmetric 
component and $D_a$ converges to a constant different from zero at large times.
Of course, this constant depends on the number of walkers $M$.
This dependence can be easily found by using the central limit theorem as before
\begin{equation}
D_a -D_e = \text{cov}( \bar{p}^t, \bar{W}^t) = K \frac{1}{M}.
\end{equation}
Finally, we have just proved that, when the guiding function is not symmetric, 
the asymptotic behaviour of the denominator is $D_a \propto \frac{1}{M}$.
This important result is in agreement with the numerical data shown in figure (\ref{fig3}).
Using exactly the same arguments, the asymptotic behaviour (large $M$, large $t$) 
of the difference of the two numerators in the R.H.S. of expression (\ref{eamcov}) 
is found to be the same as $D_a-D_e$
\begin{equation}
N_a-N_e = \text{cov}( \bar{h}^t, \bar{W}^t) \propto  D_a -D_e.
\end{equation}

Now, we are ready to write down the expression of the systematic population control error, 
$\Delta E_F^M$. For $M$ large enough, $\Delta E_F^M$ is well approximated by its first-order 
contribution in the $\frac{1}{M}$ expansion. Here also, we need to 
distinguish between the nature of the guiding function

(i) If $\psi_G$ is symmetric one easily obtains
\begin{equation}
\Delta E_F^M \propto \frac{1}{M}
\label{deamsym}
\end{equation}
(ii) If $\psi_G$ has an antisymmetric component 
\begin{equation}
\Delta E_F^M \propto \frac{1}{M} e^{\tilde{\Delta}_{B-F} t}
\label{deamnonsym}
\end{equation}

Let us now evaluate the total systematic error resulting from using a finite time $t$ and 
a finite population size $M$
\begin{equation}
\Delta E_0^F = \Delta E_F^M + \Delta E_F^t
\label{deltaeatot}
\end{equation}
where $\Delta E_F^t$, the systematic error coming from a finite simulation time, is given by 
Eq.(\ref{deltaeat}) and $\Delta E_F^M$ is the error just discussed.  
The strategy consists in determining, for a given systematic error $\Delta E_0^F \sim \epsilon$,
what is the time $t$ and the number of walkers $M$ one should consider.
The condition for the total systematic error to be of order $\epsilon$ is that both terms in 
(\ref{deltaeatot}) are also of order $\epsilon$
\begin{eqnarray}
\Delta E_F^t & \sim & \epsilon \\
\label{eatpsi}
\Delta E_F^M & \sim & \epsilon.
\label{eampsi}
\end{eqnarray} 
This is true because no error compensation are present, $\Delta E_F^t$ and $\Delta E_F^M$ 
being generally of the same sign (both positive). Our numerical results on 
the toy model give an illustration of this property. 
>From both equations (\ref{eatpsi}) and (\ref{deltaeat}) one can deduce the simulation time 
corresponding to such a systematic error
\begin{equation}
t \sim -\frac{\ln \epsilon}{\Delta_F}.
\label{teps}
\end{equation}
In other words, to obtain an error of order $\epsilon$ it is sufficient to stop the simulation 
at a time $t$ of order (\ref{teps}).
Now, let us come to the number of walkers $M$ needed.
If $\psi_G$ is symmetric, we already know from expression (\ref{deamsym}) that 
the systematic error does not depend on the projection time and that the number 
of walkers $M$ and the systematic  error $\epsilon$ are related as follows
\begin{equation}
M \propto \frac{1}{\epsilon}.
\end{equation}
If $\psi_G$ is not symmetric, the equation (\ref{eampsi}) can be easily solved.
Replacing in Eq.(\ref{eampsi}), $\Delta E_F^M$ by its expression (\ref{deamnonsym}) 
and using the relation (\ref{teps}) one finally finds the number of walkers required 
to obtain a systematic error $\epsilon$.
\begin{equation}
M \propto \epsilon^{-\frac{\tilde{\Delta}_{B-F}}{\Delta_F}-1}.
\label{Meps}
\end{equation}
Let us make some important comments. First, note that in this formula the dependence on 
the guiding function is not included in the exponent, only in the prefactor.
Second, this formula shows that the FMC algorithm reduces the systematic error by lowering 
the exponent. As already mentioned, the gap is equal to $E_0^F-E_0^B$ in a standard 
release node method and equal to $E_0^F-\tilde{E}_0^B <E_0^F-E_0^B$ in FMC.
Third, in the zero-limit gap, the $\frac{1}{M}$ behaviour of the systematic error is recovered.
This formula shows that the number of walkers needed for a given accuracy, $\epsilon$, 
grows exponentially with respect to the number of electrons. Indeed, although 
the gap is indeed reduced by FMC, there is no reason not to believe that it will still be
proportional to the number of electrons. In consequence, the ``sign problem'' fully remains in FMC.

Finally, let us write the systematic error as a function of the finite population $M$ by 
inverting the preceding equation (\ref{Meps})
\begin{equation}
\epsilon \propto M^{-\gamma}
\end{equation} 
where 
\begin{equation}
\gamma \equiv
{\frac{\Delta_F}{\tilde{\Delta}_{B-F}+ \Delta_F}}
\end{equation}
This latter equation shows very clearly the respective role played by the Fermi gap, $ \Delta_F$,
and the reduced Bose-Fermi gap, $\tilde{\Delta}_{B-F}$.
%\bibliographystyle{physrev}
%\bibliography{Assaraf}

\newpage
\begin{table}[htp]
\noindent
\caption{Average meeting times $\frac{\langle T\rangle}{N}$ for the correlated and
uncorrelated cases in the non-symmetric case $c=4$, Eq.(\ref{psigpm}). In this example,
$x_{max}=3$, Eq.(\ref{xmax}), and $\tau =0.9\tau_{max}$,
where $\tau_{max}$ is the maximal time-step defined in Eq.(\ref{taumax}).}
\begin{tabular}{|c|c|c|}
\hline
Linear size $N$ & $\frac{\langle T\rangle}{N}$ Uncorr. & $\frac{\langle T\rangle}{N}$ Corr. \\
\hline
\hline
$N=3 $     & 2152(23)  & 134(1)    \\
$N=5 $     & 2162(20)  & 92(1)     \\
$N=7 $     & 2234(26)  & 75(1)     \\
$N=9 $     & 2834(27)  & 76(1)     \\
$N=11$     & 3546(38)  & 82(1)     \\
$N=13$     & 4214(41)  & 88(1)     \\
$N=15$     &           & 94(1)     \\
$N=17$     &           & 102(1)    \\
\hline
\hline
\end{tabular}
\raggedright
\label{table1}
\end{table}

\begin{table}[htp]
\noindent
\caption{Average meeting times $\frac{\langle T\rangle}{N}$ for the correlated and
uncorrelated cases in the symmetric case ($c=0$, symmetric guiding function), Eq.(\ref{psigpm}).
In this example, $x_{max}=3$,
Eq.(\ref{xmax}), and $\tau =0.9\tau_{max}$,
where $\tau_{max}$ is the maximal time-step defined in Eq.(\ref{taumax}).}
\begin{tabular}{|c|c|c|}
\hline
Linear size $N$ & $\frac{\langle T\rangle}{N}$ Uncorr. & $\frac{\langle T\rangle}{N}$ Corr. \\
\hline
\hline
$N=3 $     & 3.32(2) & 1.634(9) \\
$N=5 $     & 5.64(6) & 2.27(1)  \\
$N=7 $     & 7.6(1)  & 2.65(1.6) \\
$N=9 $     & 10.3(1) & 3.32(2.5) \\
$N=11$     & 12.95(8)& 4.18(4)   \\
$N=13$     & 15.7(1) & 4.78(4)   \\
$N=15$     & 18.5(2) & 5.52(6)   \\
$N=17$     & 21.6(2) & 6.26(7)   \\
\hline \hline
\end{tabular}
\raggedright
\label{table2}
\end{table}

\begin{table}[htp]
\noindent
\caption{$N=3$. Reduced Bose-Fermi gap $\tilde{\Delta}_{B-F}$, Eq.(\ref{gapdelta}), 
with or without correlation for different values of $c$. 
The average meeting times are also indicated in parentheses. The bare 
Bose-Fermi gap, ${\Delta_{B-F}}$, is $\sim 0.7695$
(here, $x_{max}=3.$ and $\tau=0.09\tau_{max}$).
}
\begin{tabular}{|c|c|c|}
\hline
Value of c &
Correlated process &
Uncorrelated process\\
\hline
\hline
$c=0\;\;\;\;$ & $\tilde{\Delta}_{B-F}= 0.0366\;\;\;\;$ [$\frac{\langle T\rangle}{N}$= 1.634(9)]& 
$\tilde{\Delta}_{B-F}= 0.1629\
;\;$
[$\frac{\langle T\rangle}{N}$= 3.32(2)]$\;\;\;\;$\\
$c=1\;\;\;\;$ & $\tilde{\Delta}_{B-F}= 0.0917\;\;\;\;$ [$\frac{\langle T\rangle}{N}$= 6.37(7)] & 
$\tilde{\Delta}_{B-F}= 0.2540\
;\;$
[$\frac{\langle T\rangle}{N}$= 16.5(2)]$\;\;\;\;$\\
$c=2\;\;\;\;$ & $\tilde{\Delta}_{B-F}= 0.1336\;\;\;\;$ [$\frac{\langle T\rangle}{N}$= 24.6(2)] & 
$\tilde{\Delta}_{B-F}= 0.2277\
;\;$
[$\frac{\langle T\rangle}{N}$= 130.5(5)]$\;\;\;\;$\\
$c=3\;\;\;\;$ & $\tilde{\Delta}_{B-F}= 0.1026\;\;\;\;$ [$\frac{\langle T\rangle}{N}$= 62.9(3)] & 
$\tilde{\Delta}_{B-F}= 0.1981\
;\;$
[$\frac{\langle T\rangle}{N}$= 627(3) ]$\;\;\;\;$\\
$c=4\;\;\;\;$ & $\tilde{\Delta}_{B-F}= 0.1092\;\;\;\;$ [$\frac{\langle T\rangle}{N}$= 134(1)]  & 
$\tilde{\Delta}_{B-F}= 0.1787\
;\;$
[$\frac{\langle T\rangle}{N}$= 2152(23)]$\;\;\;\;$\\
\hline \hline
\end{tabular}
\raggedright
\label{gapf(c)}
\end{table}

\begin{table}[htp]
\noindent
\caption{Comparison between the reduced Bose-Fermi gap, $\tilde{\Delta}_{B-F}$, and 
the bare Bose-Fermi gap, ${\Delta}_{B-F}$, in the symmetric 
case ($c=0$, symmetric guiding function) as a function of $N$}
\begin{tabular}{|c|c|c|c|}
\hline
Value of $N$ & $\tilde{\Delta}_{B-F}$ & ${\Delta_{B-F}}$ & Gap ratio \\
\hline
\hline
$N=3$      & $\tilde{\Delta}_{B-F}= 0.0366$ & ${\Delta_{B-F}}= 0.7695$ & 
$\frac{\tilde{\Delta}_{B-F}}{{\Delta_{B-F}}}=0.0476$   \\
$N=5$      & $\tilde{\Delta}_{B-F}= 0.0516$ & ${\Delta_{B-F}}= 1.0195$ & 
$\frac{\tilde{\Delta}_{B-F}}{{\Delta_{B-F}}}=0.0506$   \\
$N=7$      & $\tilde{\Delta}_{B-F}= 0.0577$ & ${\Delta_{B-F}}= 1.1782$ & 
$\frac{\tilde{\Delta}_{B-F}}{{\Delta_{B-F}}}=0.0490$   \\
\hline \hline
\end{tabular}
\raggedright
\label{gapf(N)}
\end{table}

\begin{center}
\begin{figure}[htp]
\includegraphics[height=12cm,width=8cm,angle=-90]{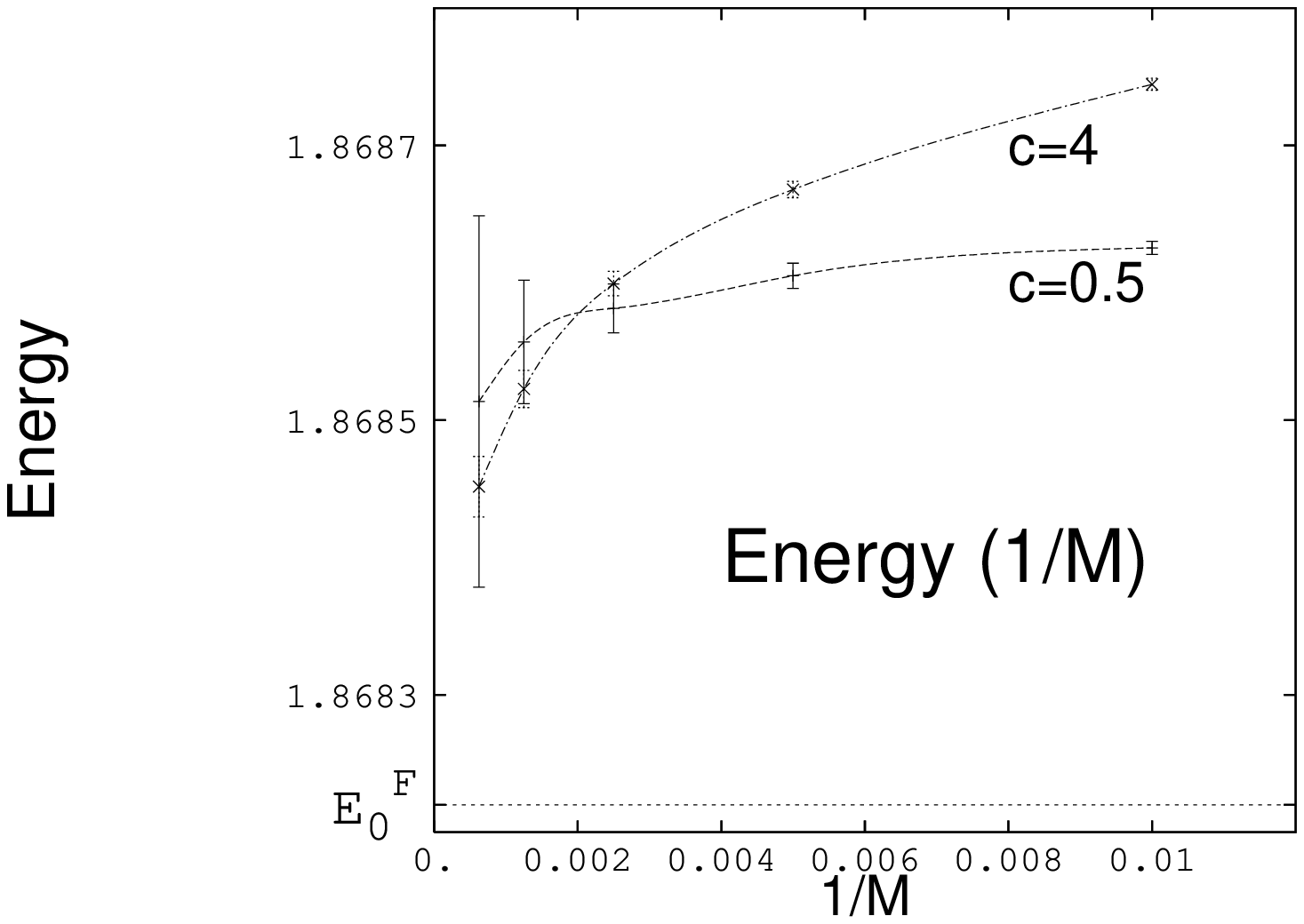}
\caption{$N=3$ Energy, Eq.(\ref{energyest}), as a function of $1/M$ for the $c=0$ and 
$c=4$ cases. Exact energy: $E_0^F$=1.86822...(horizontal solid line)}
\label{fig4}
\end{figure}
\end{center}
\end{document}